\documentclass[12pt,preprint]{aastex}

\slugcomment{Submitted to ApJ}

\shorttitle{Rotation and activity of pre-main-sequence stars}
\shortauthors{Scholz et al.}

\begin{document}
\bibliographystyle{apj}


\title{Rotation and activity of pre-main-sequence stars}


\author{Alexander Scholz} 
\affil{SUPA, School of Physics \& Astronomy, University of St. Andrews, North Haugh, St. Andrews, 
Fife, KY16 9SS, Scotland, United Kingdom\footnote{Former address: Department of Astronomy \& Astrophysics, 
University of Toronto, 50 St. George Street, Toronto, Ontario M5S\,3H4, Canada}}
\email{as110@st-andrews.ac.uk}
\author{Jaime Coffey} 
\affil{Department of Physics \& Astronomy, University of British Columbia, 6224 Agricultural Road, 
Vancouver, B.C. V6T 1Z1} 
\author{Alexis Brandeker, Ray Jayawardhana}
\affil{Department of Astronomy \& Astrophysics, University of Toronto,
    50 St. George Street, Toronto, Ontario M5S\,3H4, Canada}

\begin{abstract}
Rotation and activity are key parameters in stellar evolution and can be used to 
probe basic stellar physics. Here we present a study of rotation (measured as projected
rotational velocity $v\sin i$) and chromospheric activity (measured as H$\alpha$ equivalent
width) based on an extensive set of high-resolution optical spectra obtained with the MIKE 
instrument on the 
6.5\,m Magellan Clay telescope. Our targets are 74 F--M dwarfs in the young stellar associations
$\eta$~Chamaeleontis, TW Hydrae, $\beta$~Pictoris, and Tucana-Horologium,
spanning ages from 6 to 30\,Myr. While the H$\alpha$ equivalent widths for most F and G stars are 
consistent
with pure photospheric absorption, most K and M stars show measurable chromospheric emission.
By comparing H$\alpha$ equivalent width in our sample to results in the literature, we see a 
clear evolutionary 
sequence: Chromospheric activity declines steadily from the T~Tauri phase to the main sequence. 
Using activity as an age indicator, we find a plausible age range for the Tuc-Hor association of 
10--40\,Myr. Between 5 and 30\,Myr, we do not see evidence for rotational braking in the total sample, 
thus angular momentum is conserved, in contrast to younger stars. This difference indicates a 
change in the rotational regulation at $\sim$5--10\,Myr, possibly because disk braking cannot operate 
longer than typical disk lifetimes, allowing the objects to spin up. On timescales of 
$\sim 100$\,Myr there is some evidence for weak rotational braking, possibly due to stellar winds. 
The rotation-activity relation is flat in our sample; in contrast to main-sequence stars, there is 
no linear correlation for slow rotators. We argue that this is because young stars generate 
their magnetic fields in a fundamentally different way from main-sequence stars, and not just the 
result of a saturated solar-type dynamo. By comparing our rotational velocities with 
published rotation periods for a subset of stars, we determine ages of $13^{+7}_{-6}$\,Myr and 
$9^{+8}_{-2}$\,Myr for the $\eta$\,Cha and TWA associations, respectively, consistent with previous 
estimates. Thus we conclude that stellar radii from evolutionary models 
by \citet{1998A&A...337..403B} are in agreement with the observed radii within $\pm 15$\%.
\end{abstract}


\section{Introduction}
\label{intro}

About a decade ago, only a few nearby low-mass stars with ages between 5 and 
30\,Myr were known. Therefore, our knowledge about stellar evolution 
in this pre-main-sequence (and post-T~Tauri) phase depended largely on interpolating between 
the well-studied clusters at ages $<5$\,Myr, e.g.\ the Orion nebula cluster (ONC), IC\,348, and NGC\,2264, and 
benchmark zero age main sequence (ZAMS) clusters with ages between 30 and 100\,Myr, e.g. 
the Pleiades, IC\,2391, and $\alpha$\,Per. In the last ten years, however, several nearby 
and young stellar associations have been discovered, which provide us with the target 
samples for in-depth studies of the stellar properties in this critical age range. Most
objects in these associations are spread over large areas of the sky, and have been identified 
primarily based on satellite all-sky survey data from ROSAT, IRAS, or Hipparcos 
\citep[e.g.][]{1997Sci...277...67K,1999ApJ...516L..77M,1999ApJ...520L.123B,2000ApJ...535..959Z,
2000AJ....120.1410T,2001ApJ...562L..87Z}.

This paper concentrates on the $\eta$~Chamaeleontis cluster (hereafter $\eta$\,Cha, age 6\,Myr), 
the TW Hydra (TWA, 8\,Myr), $\beta$~Pictoris Moving Group (BPMG, 12\,Myr), and Tucana-Horologium 
(TH, $\sim 30$\,Myr) associations \citep[see][for a review]{2004ARA&A..42..685Z}. As these 
associations are all located within 100\,pc of the sun, and for all of them, a significant 
population of low-mass stars has been identified within the past few years, they are ideal 
for probing stellar evolution from the T~Tauri phase to the main sequence. Two stellar 
properties that are believed to undergo significant changes in this age range are 
magnetic activity and rotation, which are the main focus of this paper.

Magnetic activity is a collective term for all phenomena caused by the operation
of the stellar dynamo, including photospheric spots, chromospheric flares and plages, 
and coronal emission. It is observed mainly in the X-rays in the case of coronal 
activity and in optical emission lines for chromospheric processes. According to our 
current understanding, main-sequence, sun-like stars generate large-scale magnetic fields 
by means of an $\alpha\Omega$-type dynamo, which operates in the transition layer between
convective zone and radiative core \citep[see][and references therein]{2000ssma.book.....S}. 
This type of dynamo is driven by the rotation of the star, and hence a strong
rotation-activity correlation is observed for solar-type stars with ages $\ga$100 Myr,
initially found by \citet{1972ApJ...171..565S} and subsequently confirmed by numerous studies
\citep[e.g.][]{1996ApJS..106..489P,2000AJ....119.1303T,2001ApJ...561.1095B}. The correlation, 
observed both with coronal and chromospheric activity indicators, is usually seen
as a linear increase of activity indicators with increasing rotation rate, followed 
by a saturation of the activity at high rotation rates. Young T~Tauri stars, however,
are fully convective, and thus cannot harbor a solar-type dynamo. As they evolve to
the main-sequence, a change in the magnetic field generation and, as a consequence, in 
the magnetic activity properties and their connection to stellar rotation can be 
expected.

The rotation of stars themselves changes critically in the pre-main-sequence phase. As the
stars contract hydrostatically to the ZAMS, their rotation rates increase as a 
consequence of angular momentum conservation. In the first few Myrs of their evolution,
the stars are believed to lose significant angular momentum due to magnetic interaction with 
their circumstellar environments \citep[see][for a review]{2006astro.ph..3673H}, either by 
magnetic coupling between star and disk -- a process often referred to as `disk locking', 
see for example \citet{1991ApJ...370L..39K,1993AJ....106..372E} -- or by an accretion powered 
stellar wind \citep{2005ApJ...632L.135M}. Once the disk is gone and the accretion
has stopped, angular momentum removal is mainly controlled by stellar winds generated
by magnetic activity. On the main-sequence, rotational braking by winds leads to a decline 
of both rotation and potentially (via a rotationally driven dynamo, see above) magnetic 
activity \citep{1972ApJ...171..565S,2001ApJ...561.1095B}. In summary, observational studies 
of magnetic activity, rotation, and their interrelationship can probe fundamental changes 
in stellar physics occurring between $\sim 1$ and 100\,Myr.

In this paper we present a study of rotation and activity in young stars with ages between 
6 and 30 Myr, based on an extensive set of high-resolution spectra. By comparing the results 
from the associations listed above with literature data for younger and older clusters, we 
investigate the pre-main-sequence evolution of these properties. For the first time, we 
analyze a comprehensive set of H$\alpha$ line measurements together with spectroscopic 
rotational velocities $v\sin i$ for a large sample of objects in our four target regions. 
In some sense, this is complementary to the work by \citet{2004AJ....128.1812D}, who 
published a rotation/activity study for TWA, BPMG, and TH based on $v\sin i$ and X-ray data.

In a previous paper, we investigated disk accretion for these associations based on the same
multi-epoch dataset \citep{2006ApJ...648.1206J}. Our cleaned target sample for the current
paper is introduced in \S\ref{tar}. In \S\ref{obs}, we present our observations, the data reduction 
and spectral analysis, as well as a detailed assessment of the measurement uncertainties. Subsequently, 
we investigate H$\alpha$ emission (\S\ref{act}) and rotational velocities (\S\ref{rot}) as a function 
of age and mass. In \S\ref{rotact} we focus on the rotation/activity connection. By comparing $v\sin i$ 
with previously measured rotation periods, we obtain constraints on stellar radii for a subset of the 
targets (\S\ref{periods}). We summarize our results in \S\ref{conc}.

\section{Target selection and properties}
\label{tar}

This study is based on multi-epoch spectra of about 100 likely members of the four associations
listed in \S\ref{intro}. From the total sample, we excluded objects identified as accretors in 
\citet{2006ApJ...648.1206J}, because accretion affects emission lines and thus makes it difficult 
to assess chromospheric activity reliably. In \citet{2006ApJ...648.1206J}, the main criterion 
to distinguish between accretors and non-accretors is the linewidth of the H$\alpha$ emission 
feature. As thresholds we adopted 10\,\AA~for the equivalent width (EW) and 200\,km\,s$^{-1}$ for 
the 10\% width. We used the shape of H$\alpha$ and additional emission lines like He\,6678\,\AA~as
complementary probes for accretion. Four objects (two in $\eta$ Cha, two in TWA) are classified as 
accretors because their linewidths exceed both thresholds and they show He\,6678\,\AA~in emission.
One additional object, $\eta$ Cha 11, has very broad H$\alpha$ as well. Although the EW is below 
10\,\AA, we classified it as an accretor based on a strong red-shifted absorption feature in
H$\alpha$, a clear indication of infalling material. In total, five accreting objects were 
excluded from the rotation/activity analysis in this paper. 

Similarly, we excluded known unresolved binaries, since binarity might introduce additional 
line broadening. Many known binaries in these associations are listed by \citet{2004ARA&A..42..685Z},
a few more have been identified in separate papers, e.g.\ the close AO-resolved binary 
TWA\,5A \citep{2003AJ....126.2009B}. Additionally, the A-stars $\beta$\,Pic and TWA\,11A as 
well as the subgiant HD\,1555A were removed from the object list. The final cleaned target 
list comprises 74 stars. 

The spectral types for our sources have been collected from \citet{2004ARA&A..42..685Z} and 
\citet{2004AJ....128.1812D}. In a few cases where spectral types are missing in the 
literature, we obtained an approximate spectral type by simply comparing the appearance 
of the spectrum with other stars from our sample. For plotting purposes, we convert 
spectral types to a linear numerical scale, where one unit corresponds to one subclass, 
and zero is assigned to O0 stars. Spectral type F0 then corresponds to 30 while M0 corresponds 
to 60. 

Our targets span the spectral type range from F5 to M3. The distribution of spectral types
favours late-type objects, as the sample includes 36 M- and 19 K-stars. Spectral types are not
equally distributed in all four associations: With one exception, all objects in $\eta$ Cha and TWA 
are K/M-stars. In the two older groups BPMG and TH, however, the distribution is balanced with
more F/G-type objects. Effective temperatures have been derived by fitting models to the observed 
spectra, and the detailed results from this analysis will be published in a forthcoming paper 
(Mentuch et al., in prep.). In this paper, we make use of these $T_{\mathrm{eff}}$ to determine 
$L_{\mathrm{H}\alpha} / L_\mathrm{bol}$ in \S\ref{rotact}. 

The magnetic field generation and, as a consequence, magnetic activity and possibly rotation
depend critically on the stellar internal structure. For pre-main-sequence stars, it is particularly
important to clarify if they are still fully convective or if they already have developed to a solar-like 
structure with a radiative core and a convective envelope. As outlined in \S\ref{intro}, a radiative 
core is mandatory for the operation of a transition zone, solar-type dynamo, while in fully convective 
objects alternative dynamo mechanisms are at work. The age at which a radiative core appears is mass-dependent: 
While solar-mass stars are only fully convective for 1-3\,Myr, stars with 0.5\,$M_{\odot}$ need 10-20\,Myr 
to develop a radiative core. Objects with masses below 0.35\,$M_{\odot}$ are fully convective throughout 
their evolution. 

We used the theoretical evolutionary tracks by \citet{1997A&A...327.1039C} and 
\citet[][updated version 1998]{1994ApJS...90..467D} to assess the internal structure of our targets. 
In the three youngest associations, $\eta$ Cha, TWA, and $\beta$ Pic, all objects with masses 
$\gtrsim 0.6\,M_{\odot}$ are not anymore fully convective, according to the models, where the depth 
of the convection zone increases towards higher masses. For TH, the mass limit between radiative and 
fully convective is roughly at 0.3-0.4$\,M_{\odot}$, i.e.\ comparable to the main sequence. Converted to 
spectral types, this implies that all objects in our sample with spectral types earlier than M0-M2 already 
have a radiative core. In TH, this applies to all objects earlier than M3. All these limits are subject
to uncertainties due to model inconsistencies, age spread, and uncertainties in the mass/temperature/spectral
type conversion.

This essentially means that almost all our targets in $\eta$ Cha and TWA are still fully convective or 
just at the limit to develop a (small) radiative core, while in $\beta$ Pic, six out of 16 objects are 
clearly already in their radiative evolution. In TH, with its large fraction of K/G/F stars, almost all objects
have a substantial radiative core according to the models (i.e.\ basically a solar-like internal structure); 
only the three M3 stars may still be fully convective. We will use these considerations as a guideline when 
discussing evolution in magnetic activity and rotation.

\section{Observations, spectral analysis, uncertainties}
\label{obs}

\subsection{Observation and Data Reduction}

The observations were taken on 12 nights distributed over four observing runs between December 
2004 and July 2005, using the echelle spectrograph MIKE at the Magellan Clay 6.5\,m telescope on 
Las Campanas, Chile. In total, the dataset comprises of $\sim 650$ high-resolution spectra with each 
star from our sample being observed on average over 6 times. The availability of multi-epoch data for 
each star allows for a reliable assessment of variability in the emission lines and provides us
with high redundancy in the $v \sin i$ measurements. For more details about the observing
runs, see the log in \citet{2006ApJ...648.1206J}. In addition to our target stars, we observed 
a sample of $\sim 20$ slowly rotating standard stars covering the spectral range from F8 to M2 
from the list of \citet{2002ApJS..141..503N}. 

MIKE is a double echelle slit spectrograph, consisting of a blue and a red arm. In this
study, we only make use of the red part, with spectral coverage from 4\,900 to 9\,300\,\AA. 
With a 0\farcs35 slit width and no binning, our spectra have a resolution of $R\sim$60\,000.  
Depending on the brightness of the target, the integration time was up to 30\,min.  
To accommodate for the slanted spectra and the wavelength dependence of this tilt, we developed
a customized software package in ESO-MIDAS for data reduction. The details of this reduction
procedure will be discussed in a forthcoming paper (Brandeker et al., in prep.).  

\subsection{Spectral Analysis}
\label{spanalysis}

As an indicator of chromospheric magnetic activity, we measured the EW of 
the H$\alpha$ feature at 6\,562.8\,\AA~in the multi-epoch spectra in our sample. Depending 
on the spectral type and the level of chromospheric activity, H$\alpha$ is seen either as 
emission or absorption in our spectra. The EWs are obtained by integrating the H$\alpha$ line 
after continuum subtraction; absorption is defined as positive EW, emission as negative. The 
continuum is approximated by a linear fit to data points in small regions to the immediate right 
and left of the feature. These continuum defining regions are between 6\,\AA~and 10\,\AA~in 
width, which corresponds to 200 to 450 data points in the spectra.

We determined the projected rotational velocity $v\sin i$ by creating a template from spinning 
up a slowly rotating standard star of similar spectral type to the target star. The differences
in spectral types between target and standard were usually less than two subclasses. The templates 
are created in 1\,km\,s$^{-1}$ increments; we adopted as $v\sin i$ the value which produces the 
best $\chi^2$ fit between target and template. This procedure was carried out in three
spectral regions for every available spectrum of a given star, with the average value taken 
to be the final $v\sin i$ of the target star. The three spectral regions used for the fit have 
a width of 30\,\AA~and are centered at 5310, 6250, and 6815\,\AA. They are chosen to avoid
emission features and transition regions between different orders of the spectra. Since our 
earliest spectral type standard is an F7 star, we are unable to obtain $v\sin i$ measurements 
for stars that have spectral types significantly earlier than this. The availability
of multi-epoch spectra allows us several independent measurements per star for both H$\alpha$ EW
and $v\sin i$. The standard deviation from the average provides an estimate of the 
consistency of our results and the variability in H$\alpha$. All average H$\alpha$ EW and $v\sin i$
measurements are listed in Tables \ref{etacha}-\ref{th}.

For a subset of our objects, rotational velocities have already been published in the
literature, compiled recently by \citet[]{2004AJ....128.1812D}. We use 
these values as a control sample in the error analysis (see \S\ref{error}).

\subsection{Error Analysis}
\label{error}

\begin{figure}
\includegraphics[width=15cm]{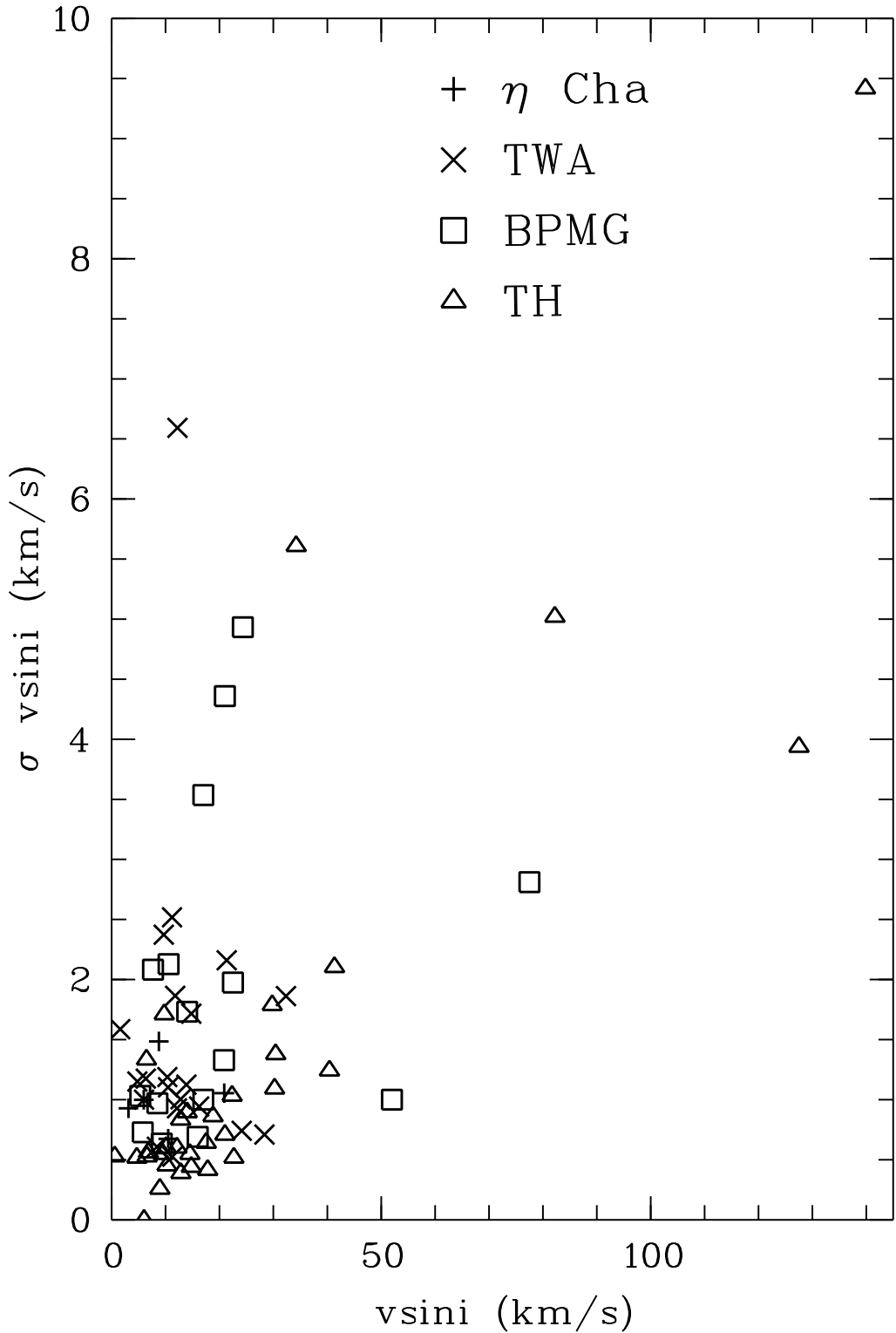}
\caption{Standard deviation of $v\sin i$ versus average $v\sin i$. Only 4 objects out of 
74 (5\%) have standard deviations greater than 5\,km\,s$^{-1}$ (see \S\ref{error}). 
\label{f1}}
\end{figure}

Strong emphasis was placed on the reliability of the rotational velocities. Particularly,
we are interested to find the lower limit in $v\sin i$ for which our results are
trustworthy. We analyze the reliability and the accuracy of our measurements in a 
variety of ways. As explained in \S\ref{spanalysis}, we measure $v\sin i$ in three 
different wavelength bins in each available spectrum of our multi-epoch dataset, giving us 
at least 3 and on average 15 independent measurements per target. The value adopted 
for the star is the average over these individual measurements. From the individual
measurements for each star we also obtain the standard deviation (of the average EW),
a measure for the spread in our datapoints for a given star, and thus a probe for the
internal consistency of the method. Fig.~\ref{f1} shows the standard 
deviation $\sigma$ versus the average value of $v\sin i$ for all objects. As can be
seen in the figure, $\sigma$ is less than 2.5\,km\,s$^{-1}$ for the majority of our
objects, confirming the reliability of our method. Only 4 out of 74 stars, or 5\%, 
have standard deviations exceeding 5\,km\,s$^{-1}$. Thus, a conservative estimate
of the internal accuracy in our $v\sin i$ values is $5$\,km\,s$^{-1}$. We adopt
this as a preliminary lower $v\sin i$ limit. The standard deviation $\sigma$ is
used in the remainder in the paper as the uncertainty in individual $v\sin i$
measurements.

\begin{figure}
\includegraphics[width=15cm]{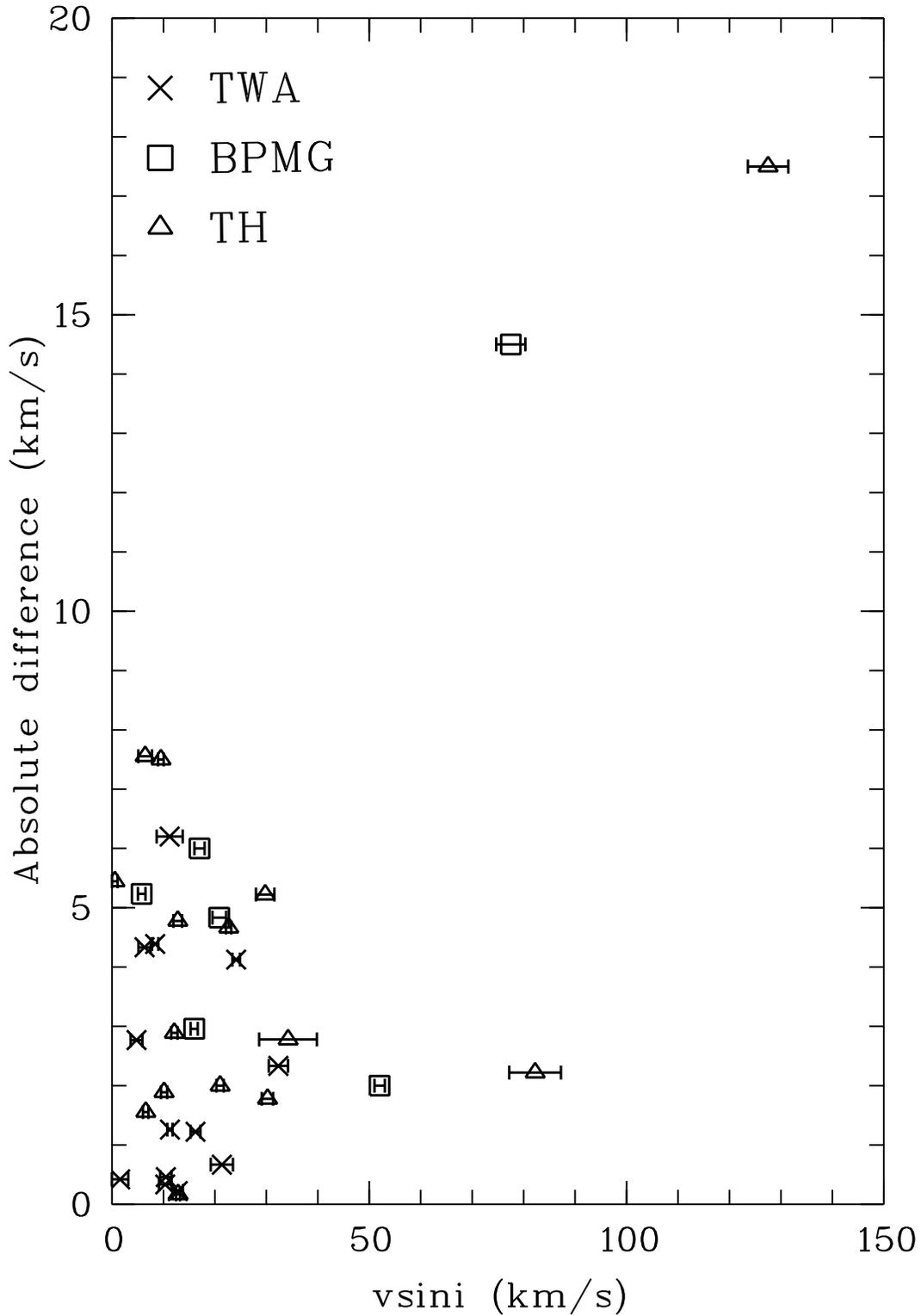}
\caption{Absolute difference between measured $v\sin i$ and literature values compiled
by \citet{2004AJ....128.1812D}. The error bars correspond to the scatter in our multi-epoch 
data. With the exception of HIP\,2729 and PZ~Tel, the deviations are not larger than 8\,km\,s$^{-1}$
\label{f2}}
\end{figure}

In the next step, we compare the measured $v\sin i$ with published values, if
available. As shown in Fig.~\ref{f2}, the majority of our values are within 8\,km\,s$^{-1}$ 
of those obtained from the literature. The two stars for which the difference between 
the measured and literature $v\sin i$ measurements exceed this amount are HIP\,2729 
and PZ~Tel, which are both fast rotators with $v\sin i$ exceeding 60\,km\,s$^{-1}$. Although 
the absolute difference between literature and our $v\sin i$ for both of these stars 
is large, the relative uncertainty is comparable to the slower rotating objects in the 
sample. Thus, the comparison with the literature confirms the validity of our
measurements. Since this comparison encompasses both the errors in our own values 
and the errors in the literature values (which are at least 3\,km\,s$^{-1}$, see
\citet{2003MNRAS.342..837R}), we find that the absolute average uncertainty in our 
$v\sin i$ measurements is comparable to the internal accuracy, which has been found 
to be 5\,km\,s$^{-1}$ (see above). Thus, we can safely trust all values down to at least 
5\,km\,s$^{-1}$.

Finally, we explore the reliability of the method by using a standard star (called A in the
following) to measure the $v\sin i$ of another standard (called B) of similar spectral type 
that has been spun-up to a determined value. That way, we produce a template with {\it known} 
rotational velocity, for which we can test the accuracy. This assumes that our
standards are slow rotators with rotational velocities well below our detection limit. 
Standard B is spun-up in steps of 1\,km\,s$^{-1}$ between 0 and 10\,km\,s$^{-1}$. We repeat 
the measurements reversing the roles of the standards, i.e.\ switching A and B. These sets 
of measurements are done for four pairs, representing cases in which we have more than one 
standard star per spectral type. This produces eight 
different pairings of standards with the same spectral type. In six of these pairings, the
measured $v\sin i$ approaches zero when the imposed $v\sin i$ approaches zero, demonstrating 
that in these cases our method produces highly accurate measurements of $v\sin i$. In two
pairings, however, the measured $v\sin i$ levels off at 5\,km\,s$^{-1}$ for imposed 
$v\sin i<3$\,km\,s$^{-1}$. The best explanation for this result is that in these cases the 
standard A has a $v\sin i$ that is 5\,km\,s$^{-1}$ greater than that of standard B, and
thus we are never able to measure values lower than 5\,km\,s$^{-1}$. This is evidence for
non-negligible rotational velocities in some of our standards, which is probably
the dominant source of uncertainty in our measurements. The test shows, however, that
these uncertainties are unlikely to exceed 5\,km\,s$^{-1}$, thus confirming the lower
reliability limit for our method. In the following, we treat all $v \sin i$ values lower
than 5\,km\,s$^{-1}$ as upper limits. We note that this affects only five objects.

We estimate the uncertainty in the measurements of H$\alpha$ equivalent width to be on average 
0.2\,\AA. This estimate was arrived at by determining the average value of equivalent width 
obtained from regions of the continuum close to H$\alpha$ that do not contain visible emission 
or absorption features. As can be seen in Tables \ref{etacha}--\ref{th}, the standard deviation
$\sigma$ in the average H$\alpha$ EW, calculated over the multi-epoch data, is in many cases clearly
higher than the measurement error, indicating significant variability (see \S\ref{var}).

\section{Magnetic activity}
\label{act}

Magnetic activity and its interplay with rotation is a complex problem, as it depends on
stellar age, mass, interior structure, and possibly interactions with disks in the early
evolution. To disentangle the different processes, we will start in this section by 
analyzing H$\alpha$ EW as a function of age and mass, We note that many of our targets have 
been identified in X-ray surveys, thus the selection might be biased towards more active objects. 
Therefore, in the discussion of activity, we prefer to use criteria based on the upper limit of 
activity in our sample rather than the lower limit, since the latter one might be biased.

\subsection{Evolution of chromospheric activity}
\label{evact}

\begin{figure}
\includegraphics[width=15cm]{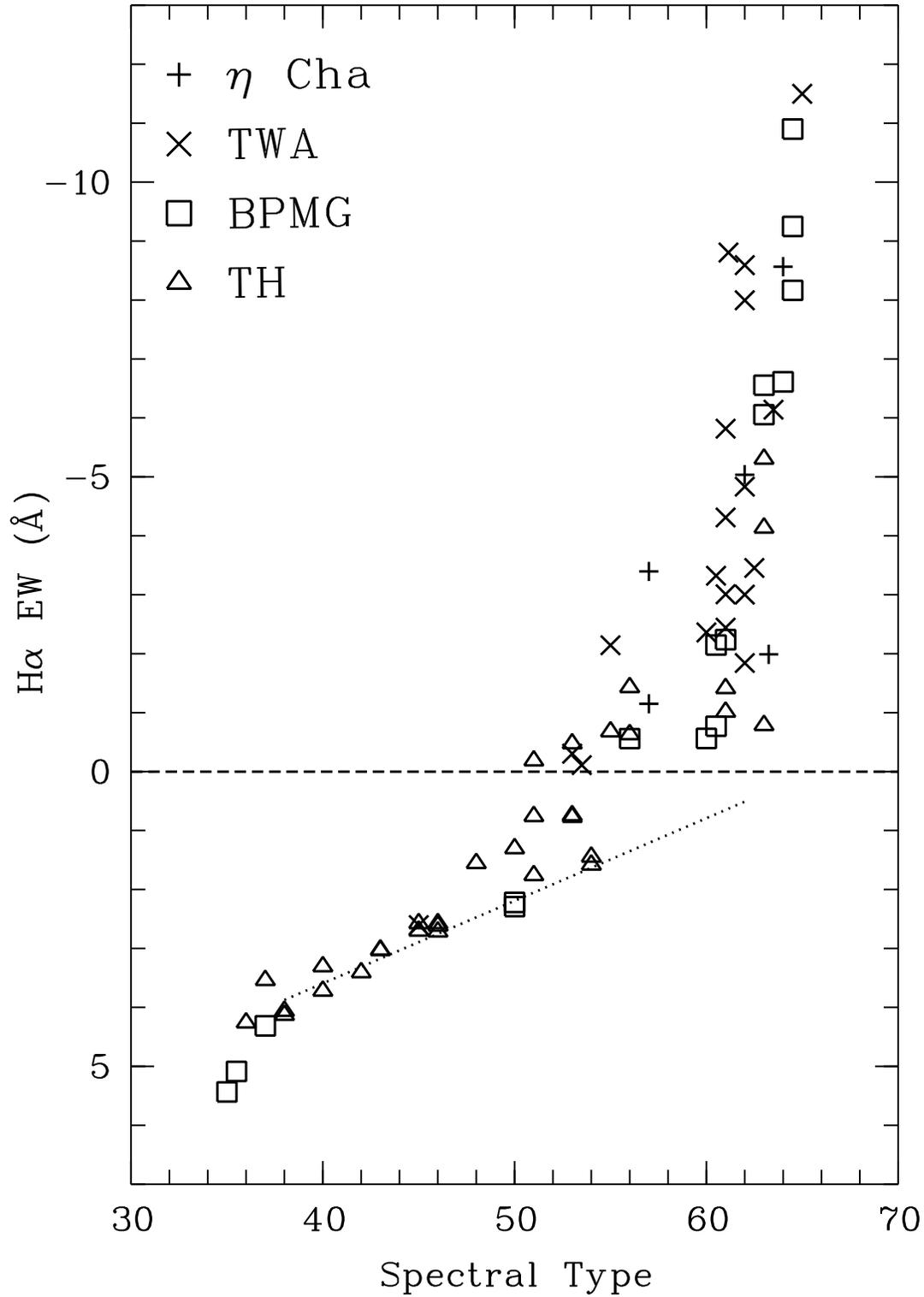}
\caption{H$\alpha$ EW as a function of spectral type. F0 corresponds to 30, M0 to 60. 
The dotted line shows the H$\alpha$ EW for non-active field stars; the dashed line
marks the zero level. \label{f4}}
\end{figure}

The H$\alpha$ feature is used routinely as an indicator of chromospheric activity, 
originating from photoionization and collisions in the hot chromosphere. A main 
tool to investigate the chromospheric activity is to plot H$\alpha$ EW vs.\ 
effective temperature, in our case represented by the spectral type. This plot 
is shown in Fig.~\ref{f4} for our young targets.

Two main features are obvious from this figure: a) There is apparently no difference in 
the distribution of datapoints for our four target regions, spanning an age range from 
6 to 30\,Myr. This is further strengthened by the fact the the distribution of EWs are 
statistically indistinguishable in the four regions. Thus, the activity levels in the four
associations are faily similar, as far as we can tell with our data (see below), although their 
ages are somewhat different. b) H$\alpha$ EWs are a strong function of spectral type. While 
mid F-type stars exhibit H$\alpha$ absorption of $\sim 5$\,\AA, the feature switches to emission 
at K2--K4 spectral types. Around M0, corresponding to masses of 0.7--0.8\,M$_{\odot}$ 
\citep{1998A&A...337..403B}, there is a clear `knee' in the distribution; at the same time, 
the spread in EW increases, and the stars reach emission levels between 0 and $-10$\,\AA~at 
early/mid M spectral types. 

The strong change of H$\alpha$ EW with spectral type does not only reflect a change in 
chromospheric activity, as the EW is additionally affected by the drop in the continuum level 
with stellar luminosity and the photospheric absorption in H$\alpha$, which is about zero for
M dwarfs and increases towards earlier spectral types. The combined effects of photospheric
continuum drop and H$\alpha$ absorption are
estimated from the H$\alpha$ EW for our standard stars, which we already used as rotational
velocity templates and which are selected to be non-active \cite[see][]{2002ApJS..141..503N}.
A linear fit to their EW as a function of spectral type is shown in Fig.~\ref{f4} as a dotted
line; the object-to-object scatter around this line is typically $\pm 0.5$\,\AA. We note that
this dashed line is consistent with the EWs of non-active stars in the Hyades and in the field 
for late K and early M spectral types \citep{1989AJ.....97..891H,1995MNRAS.272..828R}, and  
with published EW for field F and G dwarfs \citep{1964MNRAS.128..435P,1981A&AS...44..337S}.
It is also in line with theoretical predictions for H$\alpha$ EW {\it without}
chromosphere from \cite{1985ApJ...294..626C}. Thus, this line is an estimate for the pure
photospheric contribution to the H$\alpha$ EW.

As can be seen in the figure, the dotted line follows the lower envelope of the EW for our young 
target stars between F8 and M2. All EWs for spectral types earlier than G8 are in agreement with 
the dotted line and thus pure photospheric values; thus, these stars do not show measurable 
chromospheric activity. Starting at spectral type G8, the EWs measured for our targets show 
increasing excess with respect to the photospheric values, indicating a contribution from magnetic 
activity. As pointed out by \citet{1979ApJ...234..579C}, the onset of chromospheric activity will 
first tend to deepen the 
absorption feature by as much as 0.5--1\,\AA~in relatively cool and thin chromospheres, where 
line formation is dominated by photoionization and not by collisions. This effect is not seen 
in our data, no object shows significantly more absorption than the photospheric values. Thus, 
objects in transition between essentially non-active chromospheres to the collision-dominated 
regime are rare. The maximum level of magnetic activity and the fraction of active objects
increases rapidly from early K to mid M spectral types.

These results can be compared to studies of older and younger objects in a similar spectral 
range. We use three criteria: 

A) {\it The spectral type (or color) at which H$\alpha$ changes from absorption to emission.} 
This value was introduced by \citet{1999ASPC..158...63H} as an indicator of stellar age, as 
it is steadily shifting to later spectral types as the objects get older. In our sample, the 
transition is at spectral types K2--K4, but it 
is only accurately defined for objects in TH. As summarized in Fig.~5 of \citet{1999ASPC..158...63H}, 
the transition occurs at early M types in the Hyades (age 0.5--1\,Gyr), at late K types in the 
Pleiades (age 125\,Myr), and at mid K types in IC\,2602/2391 (age 30--40\,Myr). For objects in the 
1--5\,Myr age range (i.e.\ younger than our sample), the transition occurs at spectral types earlier 
than K0 \citep{2005AJ....130.1805D} -- in fact, objects without H$\alpha$ in emission are very rare 
at these ages \citep{1998ASPC..154.1772P}. Thus, the stars in TH fit nicely in the evolutionary 
sequence defined in the literature, indicating a steady decline of activity in the pre-main-sequence
evolution. Using this criterion as an age indicator, we find that ages in the TH association 
are most likely between 10 and 40\,Myr, confirming previous estimates by \citet{2000AJ....120.1410T} 
and \citet{2004ApJ...614L.125S}. 

B) {\it The fraction of K and M type objects with H$\alpha$ emission.} In our sample,
practically all K/M objects are above the photospheric values, indicating activity, which
is also the case for (non-accreting) stars younger than 6\,Myr. A close to 100\% fraction 
of active stars is also seen in pre-main-sequence clusters like IC\,2602/2391 with ages of 
30--40\,Myr \citep{1997ApJ...479..776S}. In contrast, only a small fraction of active stars 
is seen in the Hyades at an age of 0.5--1\,Gyr \citep[$\lesssim 30$\%,][]{1995MNRAS.272..828R} 
and in the old field population \citep[$\lesssim 10$\%,][]{1989AJ.....97..891H}. 
Criterion B thus confirms the drop in activity for objects older 30\,Myr, but some of this effect 
might be due to the selection bias in our sample towards highly active objects (see above).

C) {\it The maximum level of activity in M type objects.} M stars are the most active objects in our 
sample with EW ranging from zero to 11--12\,\AA. To avoid being biased too much by a few extremely 
active objects, we do not take into account the most active 10\% of the objects and thus obtain 
$\sim 9$\,\AA~in our sample. We note that this value is mostly determined by objects in the three
younger regions. For comparison, the same procedure gives $\sim 9$\,\AA~for objects
younger than 6\,Myr \citep[e.g.][]{2005AJ....130.1805D}, $\sim 6.5$\,\AA~in the Pleiades, and 
$\sim 6.0$\,\AA~in the Hyades \citep{2000AJ....119.1303T}. Again this criterion confirms a
decline of activity, which is particularly significant between 30 and 100\,Myr.
 
From all three tests, a clear evolutionary sequence is apparent. Our targets are significantly 
more active than objects in the Pleiades, Hyades, and older samples. Thus, there is a clear decline 
of chromospheric activity from 30\,Myr to 100\,Myr and beyond. This decline leads to a reduction 
of the activity level in all spectral types, which includes the complete disappearance of 
measurable chromospheric emission at earlier spectral types. As the objects age from 30\,Myr to 
1\,Gyr, the spectral range of non-active objects extends to later spectral types. Eventually, among 
evolved main-sequence stars only a fraction of M-type objects can maintain chromospheric activity. The 
best interpretation of this behavior is a mass dependence in the lifetime of active chromospheres: 
According to the numbers given above, the timescale on which the chromosphere provides significant 
H$\alpha$ emission are  $\lesssim 10$\,Myr for F and G stars, 10--100\,Myr for K stars, and 
$\ga 500$\,Myr for M stars. A similar decline of activity can probably also be seen in
the flare frequencies (see \S\ref{var})

Maybe the best approach to explain this result is based on the different internal structure of the stars
in our sample: While the F and G stars already have substantial radiative cores, most M 
stars are still fully convective. It is conceivable that once a rotationally driven $\alpha \omega$
dynamo is able to operate at the transition between convective and radiative zone, angular momentum loss
through stellar winds quickly shuts down the magnetic activity. Thus, the connection between
dynamo activity and rotation could potentially be the origin of the quick disappearance of the activity
for early-type objects. An alternative explanation for the decline of chromospheric activity 
on the pre-main-sequence might be that the change in the interior structure alters the properties 
of the magnetic surface field. This in turn might lead to a change in the dominating mechanism 
responsible for the heating of the chromosphere, which will then affect the H$\alpha$ emission.
 
As already mentioned above, we do not see significant differences between the four associations
covered with our sample. However, the spectral type coverage in the four regions is different, and thus 
it is not possible to carry out a more rigorous comparison of the four regions using the three criteria 
defined above. In TH, for example, the number of M-type objects is very low, hampering a reliable 
assessment of the fraction of active objects and their maximum activity level. On the other hand, 
the younger groups $\eta$ Cha, TWA, and BPMG lack objects to spectral types between G5 and K5, 
compromising an analysis for the onset of activity in those regions. Thus, from our data alone we 
cannot definitely rule out activity evolution between 6 and 30\,Myr. When comparing our data with
younger objects, however, we see evidence for activity evolution on this timescale, in the 
sense that the transition to emission occurs at somewhat earlier spectral types in T~Tauri
stars. Taken together, the analysis in this section indicates that chromospheric activity 
steadily declines as the stars evolve from the T~Tauri phase to the main sequence.

\subsection{Variable chromospheric activity}
\label{var}

Since we have more than one epoch for most of our targets, we are able to probe variability
in the H$\alpha$ emission. Because both photospheric H$\alpha$ absorption and bolometric 
luminosity are not expected to change significantly (i.e.\ more than a few percent) for these 
objects, variability in H$\alpha$ EW basically traces changes in the level of chromospheric 
emission. For a few objects in the youngest regions, weak levels of episodic accretion cannot 
be excluded and might contribute somewhat to the variability \citep[see][]{2006ApJ...648.1206J}.

\begin{figure}
\includegraphics[width=15cm]{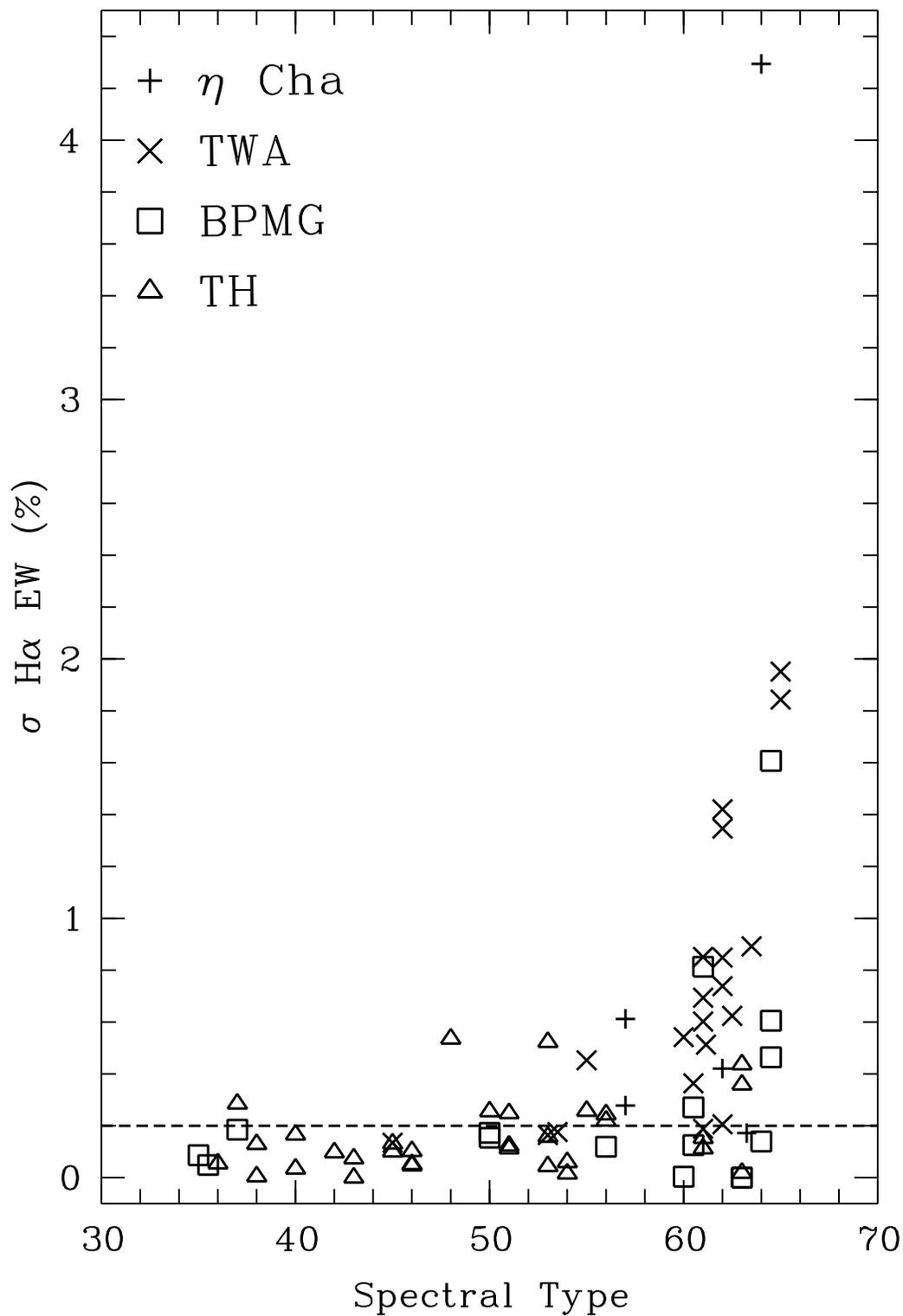}
\caption{H$\alpha$ EW standard deviation as a function of spectral type compared with 
the measurement error (dashed line). F0 corresponds to 30, M0 to 60. The datapoint for
TWA\,10 at spectral type M2.5 and H$\alpha$ EW $\sigma$ of 9.6\,\AA~is not plotted. \label{f5}}
\end{figure}

The primary estimate of variability is the standard deviations in our EW time series. In 
Fig.~\ref{f5} (left panel) we plot the absolute values of H$\alpha$ EW $\sigma$ vs.\ spectral type. 
The dashed line marks the measurement uncertainty. As can be seen from this plot, many objects with 
late spectral types show significantly higher H$\alpha$ variations than expected from
the formal error, indicating variability in activity. Interestingly, the onset of measurable
variability occurs at early K spectral types, where H$\alpha$ changes from absorption to 
emission. This confirms that the variations can indeed be attributed to chromospheric activity --
stars without measurable activity and thus only photospheric H$\alpha$ do not show variability.
The plot shows no significant difference between the four groups, indicating that the level of
variability does not strongly change between 6 and 30\,Myr.

H$\alpha$ emission originates from active regions in the chromosphere, which are typically
not uniformly distributed. Thus, one main cause of the H$\alpha$ variations is rotational 
modulation. Additionally, the light curves can be affected by flare activity and overall changes 
in the activity level, e.g.\ due to an activity cycle. Our time sampling makes it difficult
to distinguish between these three scenarios. In most cases, we have only one spectrum per
night per target; the longest time baseline is eight months. Rotational changes occur on timescales 
of the rotation periods, which are typically a few days for our targets. These changes are periodic,
but with our sparse sampling we are not able to recover the periods. General activity level changes 
are a long-term phenomenon, and thus might introduce a gradual trend in our time series. Isolated flare
events would be detectable, but only if they are clearly stronger than all other sources of 
variability. If several flares are present in our time series, it would again be difficult
to identify the source of variability.

We checked all H$\alpha$ EW time series for signs of isolated flare events. Since the typical 
flare length in the optical wavelength range is at most a few hours, a flare would appear as 
a single H$\alpha$ measurement with significantly stronger emission than all other datapoints 
in this particular time series. As a clear flare event, we accept a positive 3$\sigma$ outlier 
in the time series. It turns out that none of our objects exhibits such an event, although about 
10 of them show 2$\sigma$ outliers (for example, the active stars TWA\,10 and AU\,Mic in BPMG). Thus, 
strong, isolated flares are rare in our sample. As already mentioned, flares last typically 1--2\,h 
in the optical \citep{1999A&A...347..508G}. Given our conservative detection limit, however, we would 
detect them only in the first 10--20\,min, when their effect is most pronounced. In total, we have 
about 400 spectra, which thus cover about 100\,h. Therefore, the flare frequency derived from 
our spectra is $\lesssim 0.01\,h^{-1}$. Assuming a flare duration of $\sim 2$\,h, this corresponds
to a flare rate of $\lesssim 2$\%.

There are few reliable statistical constraints on (average) flare frequencies in the 
optical. \citet{1999A&A...347..508G} derive rates of chromospheric flares for non-accreting 
T~Tauri stars (ages $\sim 2$\,Myr) and ZAMS stars (ages $\sim 50$\,Myr) based on multi-epoch 
multi-object spectroscopy, applying a similar criterion as we have used, but for H$\beta$ instead 
of H$\alpha$. They find flare frequencies of 0.06$\,h^{-1}$ for T~Tauri and 0.006$\,h^{-1}$ for 
ZAMS stars, concluding that the average flare frequency drops by a factor of ten as the stars 
evolve from 2 to 50\,Myr. Our result of $\lesssim 0.01\,h^{-1}$ is clearly lower than the value 
derived for T~Tauri stars, which might indicate that our targets are in an intermediate 
evolutionary stage between T~Tauri phase and ZAMS in terms of their flare activity.

\section{Stellar rotation}
\label{rot}

Stellar rotation is known to be a function of mass, age, and magnetic activity. These
dependences will be discussed separately in the following subsections, with the goal
of disentangling the involved processes. 

\subsection{Rotation vs. spectral type}
\label{rotspt}

\begin{figure}
\includegraphics[width=15cm]{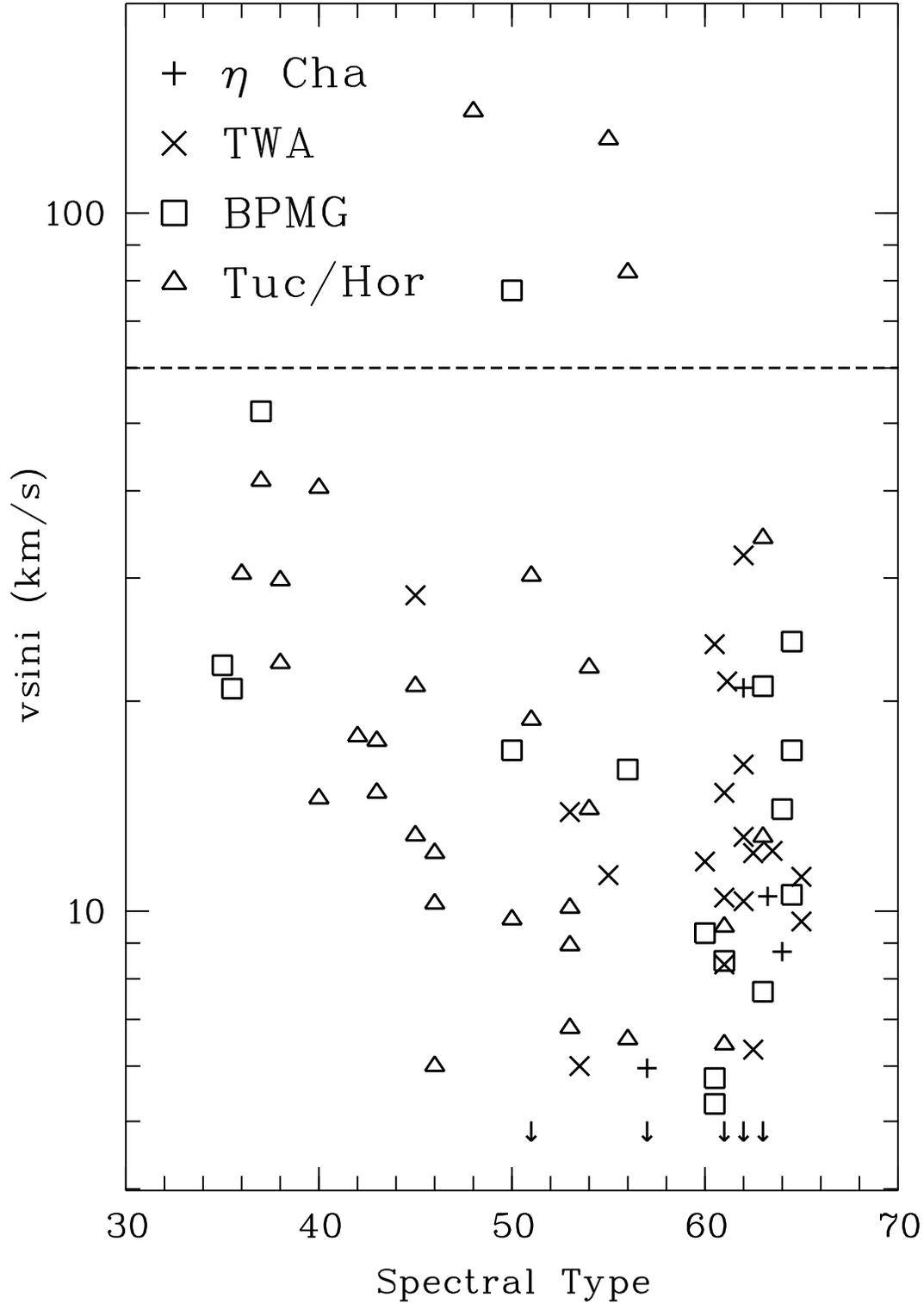}
\caption{Projected rotational velocities $v\sin i$ as a function of spectral type. 
F0 corresponds to 30, M0 to 60. The four objects above the dashed line are the so-called
ultrafast rotators in our sample.\label{f6}}
\end{figure}

Rotation is known to change as a function of stellar mass, mainly because the efficiency 
of angular momentum removal depends on magnetic activity, which in turn depends, as discussed in
\S\ref{evact}, on stellar mass. In Fig.~\ref{f6} we plot $v\sin i$ versus spectral type, 
which we use as an indicator of stellar mass. Early K spectral type roughly corresponds to 
1\,M$_{\odot}$, early M to 0.5\,M$_{\odot}$ \citep{1998A&A...337..403B}. The majority of the 
objects have rotational velocities below 60\,km\,s$^{-1}$, the four exceptions, $(5\pm 3)$\% 
of our total sample, are PZ~Tel and in BPMG and HIP\,108422, HIP\,2729, and CD-53544 
in TH. Objects with $v\sin i >60$\,km\,s$^{-1}$ are called ultrafast rotators in the following.

The overall appearance of this plot is comparable to $v\sin i$ distributions in young clusters.
In the ONC, for example, typical values for $v\sin i$ for G--M spectral types are in the range of 
12--30\,km\,s$^{-1}$, while higher mass stars tend to rotate somewhat faster \citep{2002AJ....124..546R}. 
For F--M spectral types, velocities $>$60\,km\,s$^{-1}$ are in general rare 
\citep[5--10\%,][]{2005AJ....129..363S}, consistent with our dataset. The $v\sin i$ distributions 
in ZAMS clusters like the Pleiades \citep{2000AJ....119.1303T} or IC\,2391/2602 shows the same 
phenomenological appearance. In these clusters, the number of ultrafast rotators might be somewhat 
higher ($\sim 15$\%), as expected as a consequence of pre-main-sequence contraction and thus 
rotational acceleration (see \S\ref{rotvsage}). 

For the early-type stars in our sample both the upper and the lower limit of the $v\sin i$ 
distribution decline steadily with spectral type. Excluding the ultrafast rotators, the upper limit 
decreases from $\sim 50$\,km\,s$^{-1}$ at F5 to $\sim 15$\,km\,s$^{-1}$ at K5, while the lower limit drops 
from $\sim$25\,km\,s$^{-1}$ to the detection limit of 5\,km\,s$^{-1}$ in the same spectral range. A similar 
trend is seen in the Pleiades \citep{1998A&A...335..183Q,2000AJ....119.1303T}. There are at least two
possible explanations: 

a) The timescale on which the rotation of the stars is braked as a consequence of star-disk 
interaction (see \S\ref{rotvsage}), depends on spectral type, in the sense that early type objects loose
their disks faster than later types. Evidence for mass-dependent disk lifetimes has been found recently
\citep{2006ApJ...651L..49C,2007astro.ph..1703S}, but further tests are needed to clarify the impact
on rotational evolution. 

b) The effect can also be understood as a consequence of a change in the stellar interior structure: As
already discussed in \S\ref{tar} and \S\ref{evact}, all objects earlier than M0 in our sample do have a 
radiative core and thus are able to operate a solar-type dynamo. For these objects, a deep convection zone 
enables efficient angular momentum removal due to stellar winds and/or disk-locking \citep{2000ssma.book.....S}. 
At any given age $>5$\,Myr, stars with spectral types K have deeper convection zones than F--G stars 
\citep{1994ApJS...90..467D}. Thus, as we approach later spectral types and the convection zones in the stars 
become progressively deeper, the rotational braking becomes more effective, resulting in reduced rotational 
velocities, as seen in Fig.~\ref{f6}.

\subsection{Rotational evolution in the pre-main-sequence phase}
\label{rotvsage}

\begin{figure}
\center
\includegraphics[angle=-90,width=13cm]{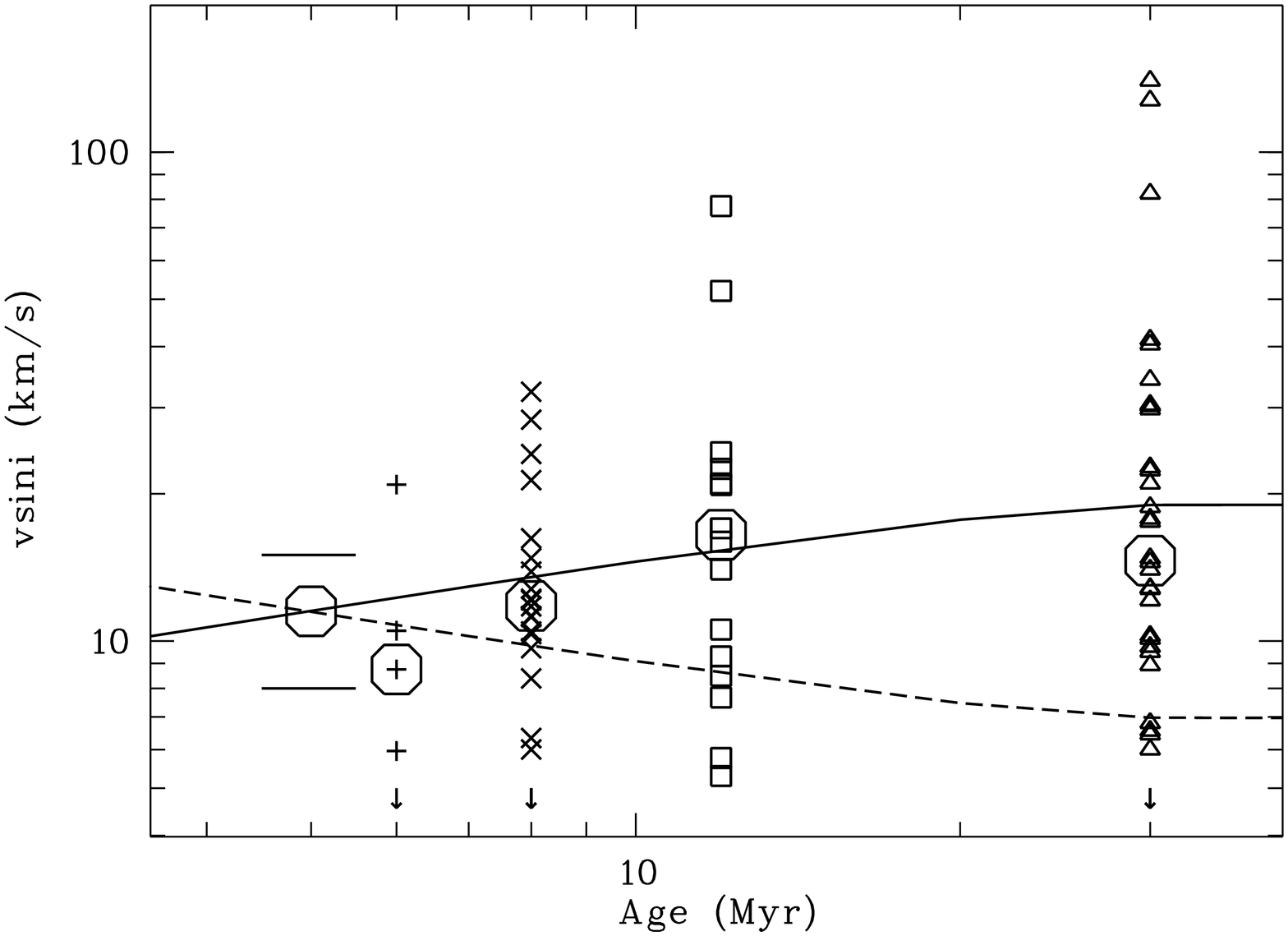}\\
\includegraphics[angle=-90,width=13cm]{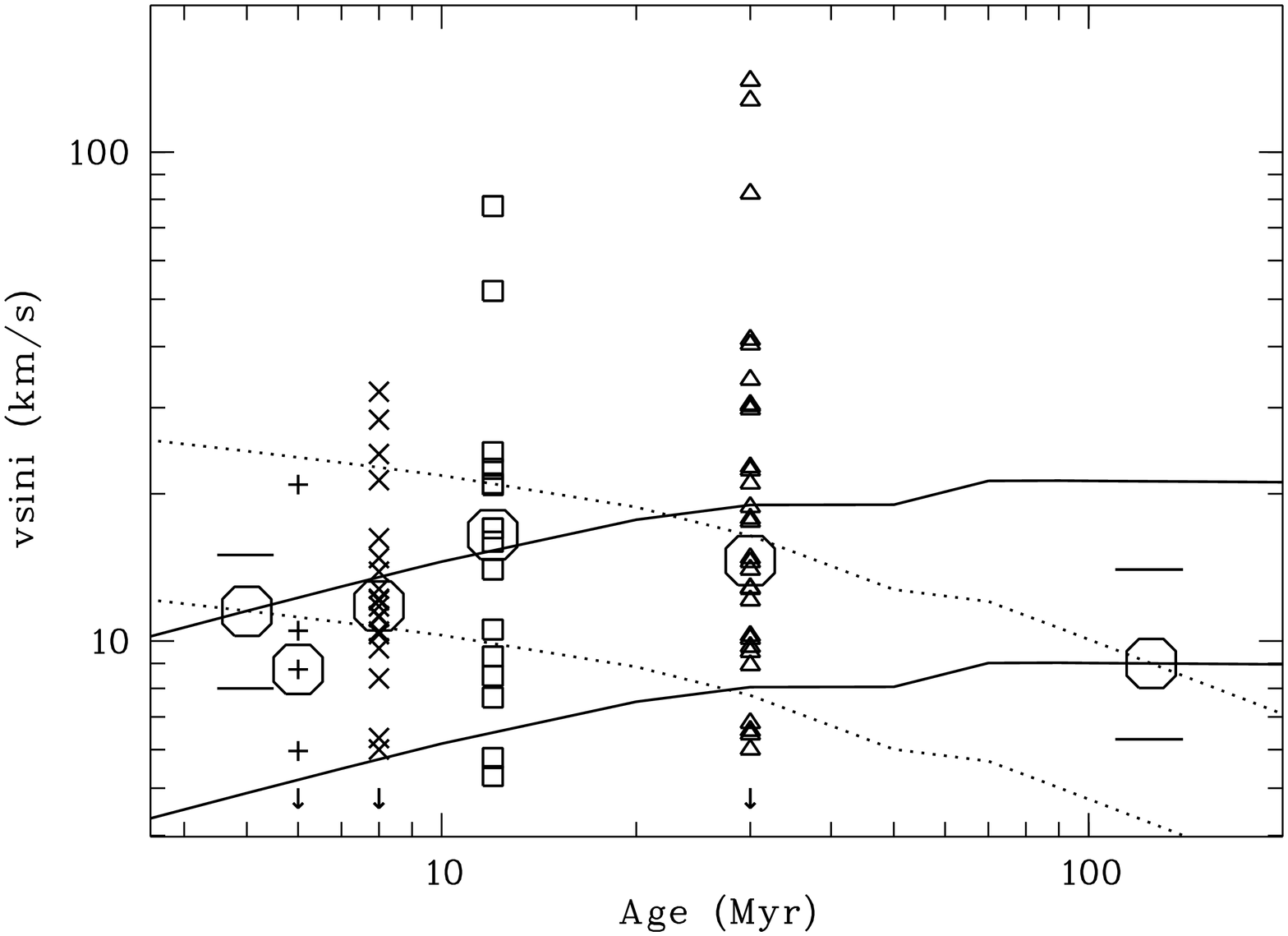}
\caption{Projected rotational velocities $v\sin i$ as a function of age, symbols as in
the previous figures. {\bf Upper panel:} Big octagons mark the median values in the four 
associations and the average at 5\,Myr (taken from \citet{2004AJ....127.1029R}). The solid 
line shows the evolution assuming constant angular momentum (i.e.\ spin-up, model A), the 
dashed line constant angular velocity (i.e.\ period locked, model B). {\bf Lower panel:} In 
addition, the median $v\sin i$ in the Pleiades is plotted (big octagon at 125\,Myr), together 
with the quartiles (horizontal lines), derived from \citet{1998A&A...335..183Q}. The solid 
lines show the rotational evolution assuming constant angular momentum (as in the upper panel,
model A), the dotted lines assume angular momentum losses following a Skumanich type law 
($v \propto t^{-1/2}$, model C).
\label{f7}}
\end{figure}

To examine the evolutionary effects more in detail, we plot $v\sin i$ vs.\ age in Fig.~\ref{f7}.
In the upper panel, we show only the four associations. As can be seen in the plot, the upper
limit in $v\sin i$ increases with age; the ultrafast rotators are only seen in older 
associations. This trend, however, might be a result of small number statistics. We compared
the distributions of $v\sin i$ using a double-sided Kolmogoroff-Smirnoff test. Specifically,
we tested the null hypothesis `the $v \sin i$ distribution in two associations is the same'.
It was found that with two exceptions all possible combinations of $\eta$ Cha, TWA, BPMG, 
and TH give likelihoods for the validity of the null hypothesis larger than 25\%. When comparing
TWA with older associations there is some marginal evidence for statistical differences,
with false alarm probabilities of 6.2\% (BPMG) and 6.3\% (TH). In general, however, the four
datasets are fairly similar. This is consistent with the results of \citet{1999AJ....117.2941S}, 
who find similar $v\sin i$ distributions for the ONC (1\,Myr) and the Pleiades (125\,Myr). 
Thus, the overall distribution of rotational velocities does not appear to change significantly 
in the pre-main-sequence phase. Please note that this does not necessarily imply consistency with 
conservation of angular momentum throughout the pre-main-sequence phase, as the objects undergo 
a strong contraction (see below for a more detailed assessment).

A large scatter of rotation rates is seen at all ages. While the projection factor $\sin i$,
age spread, and $v\sin i$ uncertainties all contribute to the scatter, the major reason for the
large spread of the distribution is probably the spread in the initial rotation periods. In clusters with 
ages of 1--2\,Myr, the periods range from fractions of a day to $\sim 20$\,d \citep{2006astro.ph..3673H}, 
corresponding to rotational velocities ranging from $<5$\,km\,s$^{-1}$ to $>100$\,km\,s$^{-1}$. Moreover, 
the $v\sin i$ distribution (as well as the distribution of $\log v\sin i$) is highly asymmetric, 
hampering a rigorous statistical analysis. To mitigate this problem when investigating the 
rotational evolution we work in the following with typical (median) $v\sin i$ values for a 
given age, rather than with individual datapoints. Please note that by averaging over the 
rotational velocities in one particular group, we loose any information about the spectral type 
dependence of the rotation, which has been discussed in \S\ref{rotspt}. 

In Fig.~\ref{f7}, we overplot the median values for each association as large octagons. We 
will compare these median $v\sin i$ with simple models for the rotational evolution. 
As starting value for the models, we used the typical $v \sin i$ of 8--15\,km\,s$^{-1}$ (average
11.5\,km\,s$^{-1}$) at $\sim 5$\,Myr given by \citet{2004AJ....127.1029R}. To take into account 
the pre-main-sequence contraction, we use radii from \citet{1997A&A...327.1039C} for a stellar 
mass of 0.8\,M$_{\odot}$, which is typical for our sample. 

In the upper panel of Fig.~\ref{f7}, we plot the expected rotational evolution for two
extreme cases, constant angular momentum with a solid line (model A) and constant angular 
velocity with a dashed line (model B). In this approach, we follow \citet{2004AJ....127.1029R}
who have done a similar comparison for stars with ages from 1--10\,Myr. In model B,
the period is constant, as expected in a scenario with ideal `disk-locking', and thus
$v\sin i \propto R$. Model A, on the other hand, shows purely the spin-up due to contraction and
thus $v\sin i \propto R^{-1}$. While both models are in good agreement with observations
until ages of $\sim 10$\,Myr, only model A is clearly consistent with the median $v\sin i$ at
30\,Myr. Model B, however, gives too low values for ages $>$10\,Myr; it truncates the $v\sin i$ 
distribution at the 20\% quartile in BPMG and at the 10\% quartile in TH. Thus, from 10 to 30\,Myr 
the objects show rotation rates rather consistent with conservation of angular momentum than with
constant rotation period. Thus, the dominating effect for the rotational evolution in this
time window is spin-up due to the pre-main-sequence contraction. This result is robust against 
uncertainties in the stellar radii, because only the ratio of radii is used in the calculation. 

In strong contrast to our finding, for ages $<5$\,Myr the rotational evolution closely follows 
the track for constant angular velocity, as concluded by \citet{2004AJ....127.1029R}. There is 
growing evidence for a strong rotational braking in the first few Myr, most likely produced by 
interaction with accretion disks \citep[e.g.][]{2002A&A...396..513H,2006ApJ...646..297R} and 
preventing the stars from spinning up by essentially locking the rotation period 
\citep[e.g.][]{2002AJ....124..546R,2002ApJ...564..877T,2005ApJ...633..967H}. Our results now 
demonstrate that while the period may be locked until ages of $\sim 5-10$\,Myr, in the following 
$\sim 20$\,Myr the stars spin up without clear evidence for rotational braking. Thus, rotational 
acceleration (measured in period) becomes significant at ages of 5--10\,Myr -- which is consistent 
with the typical lifetime of circumstellar disks \citep{2001ApJ...553L.153H}. Specifically, it
has been shown that many of the youngest stars in our sample (in $\eta$ Cha and TWA) are affected 
by inner disk clearing measured from mid-infrared excess \citep{2005ApJ...627L..57H,1999ApJ...521L.129J},
while the oldest objects (in TH) do not show any evidence for disks at mid-infrared wavelengths
\citep{2004ApJ...612..496M}. Thus, the change of the rotational regulation at 5--10\,Myr  
coincides with the disappearance of the inner disks. It has to be emphasized, however, that all 
these considerations only apply to the {\it typical} evolution. For individual objects, the 
period-locking timescale can vary by a lot -- possibly due to different disk lifetimes.

To follow the evolution to the main sequence, we compared our dataset with the rotational
velocity data in the Pleiades. In the lower panel of Fig.~\ref{f7}, we plot the median $v \sin i$
for F to M stars (large octagon) together with the quartile values (horizontal bars). These 
numbers have been taken from \citet{1998A&A...335..183Q} (their Fig.~6, averaged over all masses). 
In this plot we show for each model two evolutionary tracks, the first starts at 6\,Myr and 
calculates forward in time (as in the upper panel), the second starts at 125\,Myr and calculates 
backwards. Solid lines show again model A, i.e.\ conservation of angular momentum without any rotational 
braking. The tracks from model A are barely consistent with the observational data. When started
at 5\,Myr, the predicted median in the Pleiades is 21\,km\,s$^{-1}$ and thus too high; when started 
at 125\,Myr, they give a median of 5\,km\,s$^{-1}$ at 5\,Myr, which is clearly too low. Thus, 
rotational braking is likely involved in the evolution to the ZAMS.
 
On the main-sequence, rotation is mainly braked by angular momentum losses due to 
stellar winds, where the standard rotational braking law has been found to be $v \propto t^{-1/2}$
\citep{1972ApJ...171..565S,2001ApJ...561.1095B}. Model C, shown in dotted lines, assumes angular 
momentum losses according to the Skumanich law, again calculated in both directions. The tracks from 
model C, however, are clearly not in agreement with the observations. When calculated forward, the 
predicted median for the Pleiades is well below the detection limit. Conversely, for 5\,Myr the model 
gives an unrealistically high median. Thus, Skumanich braking appears to be too strong. We can reproduce 
the $v\sin i$ evolution either by using an exponent of $-0.1$ to $-0.3$ instead of $-0.5$ in the braking 
law, by using an exponential braking law with $v\sin i \propto \exp{(-t)}$ or by switching on the braking 
at about half way through the pre-main-sequence evolution. The latter scenario is not implausible, as most 
objects in the considered mass range develop a radiative core and thus the pre-requisite to operate a 
solar-type dynamo after about 30\,Myr, see \S\ref{intro}. 

Thus, our comparison with models gives the following results: a) On timescales of $\sim 100$\,Myr, 
weak rotational braking, possibly due to a Skumanich-type activity-rotation connection, is required to 
find a good match to the observations. b) From 5--30\,Myr the rotational evolution is fully consistent 
with angular momentum conservation; effects of possible rotational braking are too weak to affect the 
$v\sin i$ distribution significantly. Again it should be emphasised that these results do only apply to 
the total sample. In \S\ref{rotspt} we do find that rotational velocities depend on spectral type for 
objects earlier than M2. Thus, for objects with ages between 5 and 30\,Myr, stellar mass is the major 
factor which determines the rotation, rather than age.

In the previous sections we have already made connections between rotation and activity, to explain
the evolution and mass-dependence of H$\alpha$ emission and rotational velocities. The obvious next step
is to investigate directly possible correlations between rotation and activity, which is the focus of 
the next subsection. 

\subsection{The rotation-activity connection}
\label{rotact}

In order to obtain a physically meaningful picture of a possible connection between rotation and activity,
we derived relative H$\alpha$ luminosities (i.e.\ $L_{\mathrm{H}\alpha} / L_\mathrm{bol}$) from the 
measured H$\alpha$ EW. We focused on the objects with clear chromospheric H$\alpha$ emission, and 
therefore excluded stars with spectral type earlier than K2 (see \S\ref{evact}). In a first 
step, we corrected the EW for photospheric absorption, using the correlation between
photospheric H$\alpha$ absorption and spectral types derived in \S\ref{evact} from non-active
reference stars (see dotted line in Fig.~\ref{f4}). Objects with corrected EW $<0.5$\,\AA~and thus 
insignificant chromospheric emission were excluded. The continuum at the wavelength of H$\alpha$
was estimated using the STARdusty1999 model spectra, which are based on the NextGen models refreshed
with new water and TiO opacities  \citep{2000ApJ...540.1005A}. We measured the continuum flux at 
6562\,\AA~for effective temperatures ranging from 3\,000 to 5\,000\,K and $\log g = 4.0$ by approximating 
the spectrum around H$\alpha$ with a linear fit. This value was divided by the bolometric luminosity 
for the respective effective temperature. As a result, we obtain scaling factors as a function of 
effective temperature to convert the H$\alpha$ EW to $L_{\mathrm{H}\alpha} / L_\mathrm{bol}$. Please note 
that this conversion depends neither on the radii of the objects nor on the distances, which are poorly 
constrained for many of our targets. The effective temperatures for our targets will be published in a 
forthcoming paper, see \S\ref{tar}.

\begin{figure}
\includegraphics[angle=-90,width=15cm]{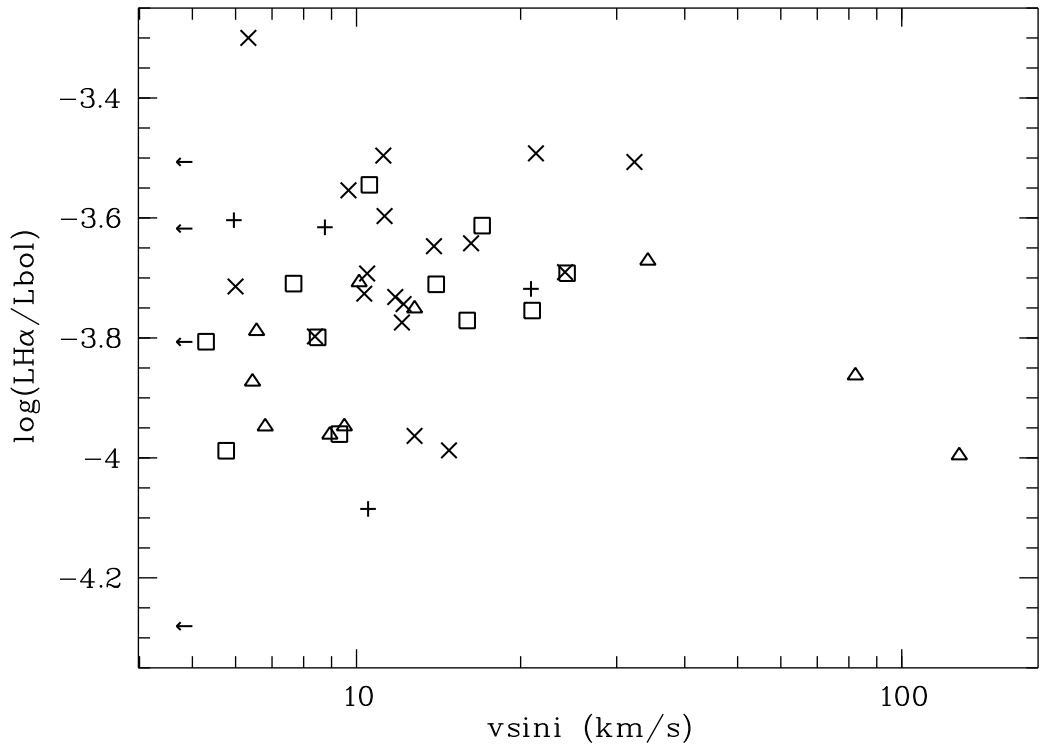}
\caption{Relative H$\alpha$ luminosities $L_{\mathrm{H}\alpha}/L_\mathrm{bol}$ as a function of
$v\sin i$. Symbols are the same as in Fig.~1--6. Plotted are only objects with significant chromospheric 
H$\alpha$ emission (which implies spectral type later or equal K2). \label{f8}}
\end{figure}

Fig.~\ref{f8} shows the relative H$\alpha$ luminosities as a function of $v\sin i$. Please note that by 
excluding non-active (earlier type) objects, the clear majority of the objects in the plot is fully convective. 
While the lower activity limit in this plot is a detection limit, the upper limit is reliably determined 
and can be compared with published samples. In our sample, we obtain $\sim 3 \times 10^{-4}$,
excluding the datapoint for TWA\,10, which possibly is affected by a flare event (see \S\ref{var}). 
For the mass range of our sample, this value is roughly comparable with the upper limit in the Pleiades 
\citep{1995MNRAS.274..869H}, but clearly higher than in the Hyades 
\citep[$\sim 1.4 \times 10^{-4}$,][]{1997ApJ...475..604S}, again indicating a decline of the general activity 
with age, as already discussed in \S\ref{evact}.

As can be seen in Fig. \ref{f8}, the upper limit of the range in activities is mostly flat. Thus, activity 
is not strongly correlated with mass for 5\,km\,s$^{-1} \le  v\sin i$. This holds even when we only consider objects with 
radiative core and thus the potential to operate a solar-type, rotationally driven dynamo. It is also important 
to note that the distribution of rotational velocities for the non-active stars (not contained in Fig.~\ref{f8}) 
is indistinguishable from the active stars; they cover the full range from $<5$ to 100\,km\,s$^{-1}$, with a 
accumulation between 10 and 20\,km\,s$^{-1}$. Moreover, among the four slowest rotators in our sample with 
$v\sin i <5$\,km\,s$^{-1}$ (shown as upper limits in Fig. \ref{f8}), there is only one object with an activity 
level significantly below the range of datapoints for the faster rotators. Thus, by and large the rotation-activity 
correlation derived from H$\alpha$ emission is flat in our sample.

These results can be compared phenomenologically with rotation-activity studies based on X-ray data. 
\citet{2004AJ....128.1812D} find that stars in TWA, BPMG, and TH are roughly comparable to T~Tauri stars 
in the ONC in terms of their X-ray properties. The activity in the ONC has been studied in detail in the 
COUP project \citep[e.g.][]{2003ApJ...582..398F,2004AJ....127.3537S}. Both in the COUP data and in the 
sample of \citet{2004AJ....128.1812D}, there is no strong correlation between $L_x / L_\mathrm{bol}$ and 
rotation period. The rotation/activity relationship appears to be flat over a wide range of periods, 
interpreted as saturation with some indication for supersaturation, i.e.a decline of activity
for the fastest rotators. This is very similar to what we observe in H$\alpha$. The two ultrafast
rotators in Fig. \ref{f8} appear to have below average activity levels, which might be interpreted as 
supersaturation. The two additional ultrafast rotators not plotted in Fig. \ref{f8} have no measurable
activity level, thus confirming this trend. However, since we have only very few datapoints at high 
rotational velocities, this should be treated with caution. Still, it is interesting to note that the 
four ultrafast rotators are objects with radiative core, maybe implying that supersaturation might be 
associated to the presence of a solar-type dynamo.

It is well established that field stars show a mostly linear relationship between rotation and 
relative X-ray luminosity \citep{2000ASPC..198..401R}. In young open clusters like IC\,2391, IC\,2602, 
and the Pleiades with ages ranging from 30 to 150\,Myr, an intermediate situation is seen, with 
many objects in the saturated regime and an additional linear part \citep[e.g.][]{1996ApJS..106..489P}.
A hint of a linear relation might also be seen in the sample of post T~Tauri Lindroos stars
with ages between 10 and 100\,Myr analyzed by \citet{2004A&A...428..953H}. Linear correlations between 
X-ray flux and rotation rate have additionally been found for young stars in Taurus 
\citep{2001A&A...377..538S}. \citet{2004AJ....127.3537S} argued that the linear part of the 
rotation/activity correlation in the ONC may be hidden in the objects for which no periods have been 
measured. However, studies of magnetic activity at very young ages are problematic, because accretion 
additionally affects both X-ray and H$\alpha$ luminosities, which in principle requires the strict 
separation of accretors from non-accretors. Our H$\alpha$ luminosity vs. $v\sin i$ plot does not reveal 
a strong indication for a linear regime in the rotation/activity relationship. The linear part cannot
be hidden at low and thus undetectable rotational velocities, as it has recently been found for field
M stars \citep{2007astro.ph..2634R}, because even the slowest rotators in our sample show the same 
range of activity levels (with one exception, see above). In summary, it is still 
not clear if the linear part of the rotation/activity correlation is already established at ages 
$<30$\,Myr.

In this context it is interesting to note that the rotation/activity relation of young
stars is similar to very low mass (VLM) objects with masses $<$0.3\,M$_{\odot}$. It is known that the 
rotational velocity at which saturation is reached drops quickly with decreasing object mass 
\citep{2003A&A...397..147P}, with the result that most VLM objects appear in the saturated regime 
\citep{1998A&A...331..581D,2003ApJ...583..451M}. As a consequence, the Skumanich type braking law
breaks down \citep{2000ApJ...534..335S}, and a weak exponential braking law is expected with 
$v_{\mathrm{rot}} \propto \exp{(-t)}$ \citep{2003ApJ...586..464B}. Such weak rotational braking
is indeed required to model the rotational evolution in the VLM regime on timescales of 
$\sim$100--1\,000\,Myr \citep{2004A&A...421..259S,2005A&A...429.1007S}.

In the canonical picture of the rotation/activity connection, the solar-type $\alpha\Omega$ dynamo 
strongly depends on rotation, causing a linear relationship at low and moderate activity levels. 
The saturation effect is usually interpreted as an activity level where the stellar
surface is covered by magnetic flux tubes and no further enhancement of activity (and
rotational braking) is possible -- thus the term `saturation'. (It is unlikely that this 
corresponds to a surface completely covered with starspots, given the fact that many `saturated' 
stars show strong photometric modulations due to rotation and thus have only a partially 
filled surface.) Both VLM objects and very young stars, however, are fully convective and 
thus cannot harbor a solar-type dynamo, which operates at the transition between convective
and radiative zone (see \S\ref{intro}). The kind of alternative dynamo that produces their 
magnetic activity, and how it depends on rotation, is still a matter of debate
\citep[see e.g.][]{1993SoPh..145..207D,2006A&A...446.1027C,2006Sci...311..633D}. But it is
at least questionable to assume that the picture of the rotation/activity connection used
for evolved solar-type stars can simply be extended to young stars and very low mass objects. 
For these types of objects `saturation' can have two meanings: a) They are saturated and do 
not follow the Skumanich law, because they rotate too fast to be in the linear regime, as it is 
assumed in the standard paradigm. b) Their `saturation' is the consequence of a magnetic field 
generation fundamentally different from solar-type stars.

In the second case, `saturation' is not merely the consequence of fast rotation, but a more
fundamental sign of a change in the magnetic field generation (and thus the word `saturation'
might be misleading). This idea has been proposed as an interpretation of rotation and activity 
data for open cluster stars by \citet{2003ApJ...586L.145B,2003ApJ...586..464B}. Basically, the Barnes
scheme suggest that the rotation/activity properties can be understood {\it only} in terms
of the magnetic field generation: Fully convective objects in the saturated or supersaturated regime 
(on the `C-sequence' in the nomenclature of \citet{2003ApJ...586L.145B}) do not harbor a solar-type 
magnetic dynamo and thus do not follow the Skumanich type rotational braking law. As the stars
evolve from the T~Tauri phase to the ZAMS (and develop a radiative core), the fraction of objects 
on the C-sequence drops quickly and reaches values $<10$\% at the age of the Hyades. While the 
quantitative predictions of the \citet{2003ApJ...586L.145B} scheme may not be convincing in all
cases, the qualitative picture is consistent with the current rotation/activity data 
for the pre-main-sequence evolution.

Many of our target stars in $\eta$\,Cha, TWA, BPMG, and TH are too old to be fully
convective (see \S\ref{tar}). Depending on their mass, they have developed
radiative cores with substantial radii already. Thus, they present an interesting test case 
for magnetic field evolution. The fact that they show rotation/activity properties similar to
younger stars (and to fully convective VLM objects) might indicate that it takes at least
30\,Myr until the solar-type dynamo dominates the magnetic activity and rotational braking.
This is supported by the weak rotational braking on the pre-main sequence found in \S\ref{rotvsage}. 
Future investigations of the magnetic field properties of pre-main-sequence stars as a function 
of age hold great potential to clarify these issues.

\section{Constraints on ages and radii: Rotation periods vs.\ $v\sin i$}
\label{periods}

By combining our measured $v\sin i$ with previously published rotational periods, we can derive 
stellar radii (times the unknown projection factor $\sin i$) as $R \sin i = (2\pi)^{-1} P v\sin i$, 
and compare these to evolutionary models, to constrain the age. This gives an age estimate independent
of other indicators such as the lithium abundance and color-magnitude diagram constraints, 
although all these estimates are dependent on the particular evolutionary model used. The 
estimated $R\sin i$ thus test the self-consistency of the models.

\begin{figure}
\includegraphics[angle=0,width=15cm]{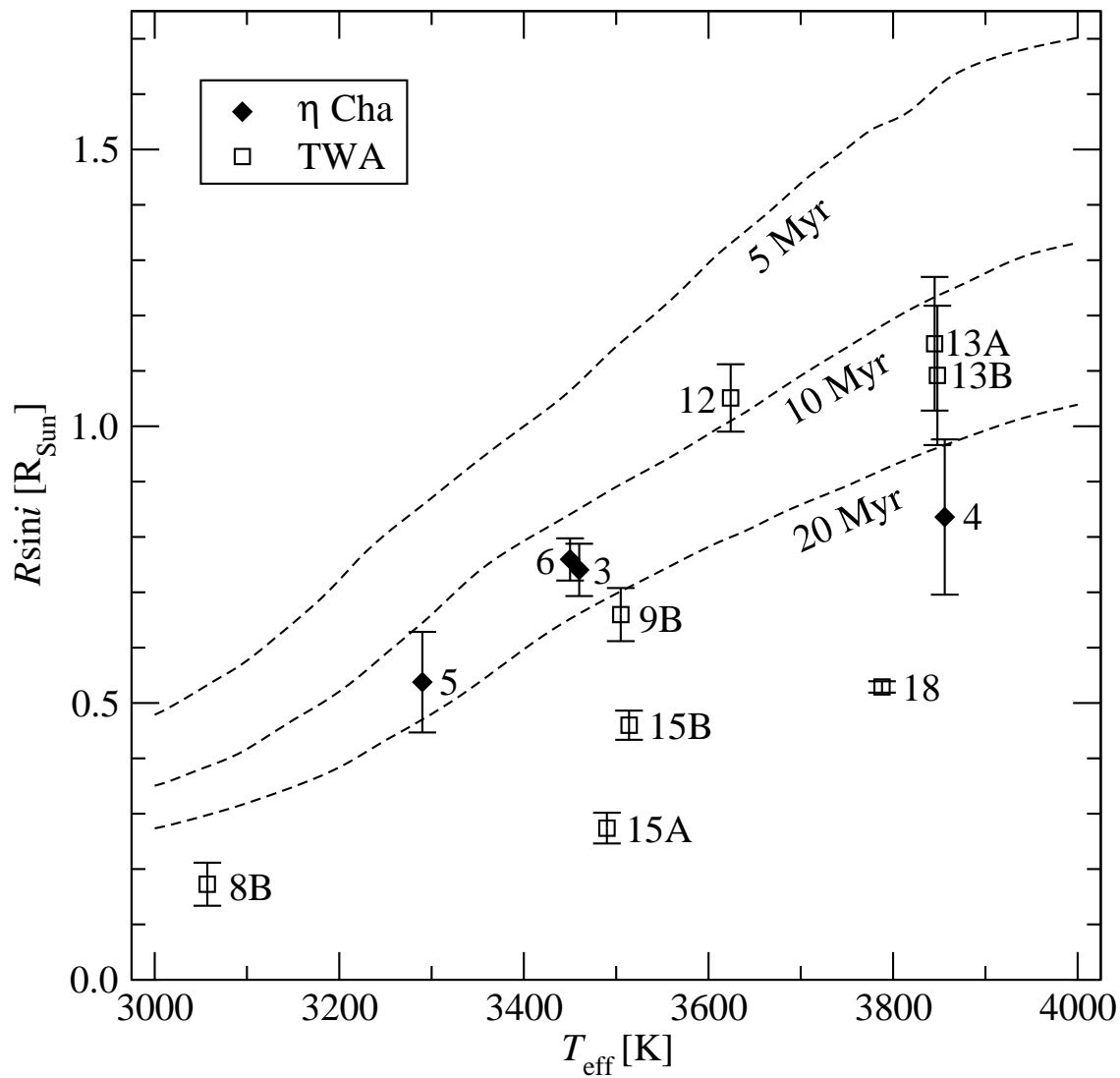}
\caption{Derived $R\sin i$ for 4 targets in $\eta$\,Cha and 8 in TWA. The dashed lines with labeled
ages are the radius isochrones from \citet{1998A&A...337..403B}, and effectively correspond to
upper limits on $R\sin i$ for objects of a given age. Even casual inspection clearly puts
the age of these association between 5 and 20 Myr. More careful analysis yields $t_{\eta\,\mathrm{Cha}} = 
13^{+7}_{-6}$\,Myr and $t_{\mathrm{TWA}} = 9^{+8}_{-2}$\,Myr for $\eta$\,Cha and TWA, 
respectively (see \S\ref{periods}).\label{f9}}
\end{figure}

Because of the unknown projection factor $\sin i$, a statistical sample is needed to derive 
the true $R$. Unfortunately, since our targets are widely distributed in the sky, monitoring 
campaigns to find photometric periods are time consuming, and only a small subset of our sample 
has photometric periods measured. In total, 16 periods have been measured in $\eta$\,Cha 
\citep{2001MNRAS.321...57L} and TWA \citep{2005MNRAS.357.1399L}. Of those, 13 are M dwarfs, out 
of which 12 have $v\sin i$ above our detection threshold. In Fig.~\ref{f9} we plot $R\sin i$ against 
effective temperature for those 12 targets, together with radius isochrones from models by 
\citet{1998A&A...337..403B}. The errors in the $R\sin i$ are entirely dominated by errors in 
$v\sin i$.

Although the statistical sample (12) may seem small, we are helped by the fact that the probability 
density distribution for the projection factor $\sin i$ of a random orientation favors close to 
edge-on geometries (see, e.g., Appendix A in \citet{2006astro.ph..8352B}): 

\begin{equation}
\label{e:sini}
f(\sin i) = \frac{\sin i}{\sqrt{1 - \sin^2 i}}.
\end{equation}

Assuming a single age and the evolutionary models by \citet{1998A&A...337..403B}, equation 
(\ref{e:sini}) can be used to find a maximum-likelihood estimate for the age. To take into account 
the estimated measurement error $\sigma$, and to mitigate the singularity at $\sin i = 1$, we assume
the measured $R\sin i$ to be an outcome of a stochastic variable 
$\mathcal{R} \in R_{\mathrm{mod}}(t,T_{\mathrm{eff}}) Y + E$, where $Y = \sin i$ is the projection 
factor distributed according to equation (\ref{e:sini}), $E$ is normally distributed with zero mean 
and variance $\sigma^2$, and $R_{\mathrm{mod}}(t,T_{\mathrm{eff}})$ is the model radius for a star 
of age $t$ and effective temperature $T_{\mathrm{eff}}$. The probability distribution of $\mathcal{R}$ 
is obtained by numerical integration,

\begin{eqnarray}
f_{\mathcal{R}}(r|R_{\mathrm{mod}},\sigma) &=& \frac{1}{R_{\mathrm{mod}}}
 \int^{\infty}_{-\infty} f_Y\left(\frac{x}{R_{\mathrm{mod}}}\right) f_E(r-x)\,\mathrm{d}x \nonumber\\
 &=& \frac{1}{R_{\mathrm{mod}}\sigma\sqrt{2\pi}} \int^{R_{\mathrm{mod}}}_0
   \frac{x \exp[-(r-x)^2/(2\sigma^2)]}{\sqrt{R_{\mathrm{mod}}^2 - x^2}}\,\mathrm{d}x, \label{e:R}
\end{eqnarray}

and the maximum likelihood by finding the maximum of the likelihood function

\begin{equation}
\mathcal{L}(t) = \sum_j \log f_{\mathcal{R}}(r_j|R_{\mathrm{mod},j},\sigma_j).
\end{equation}

To estimate conservative confidence intervals for this estimate, we integrate the probability 
density function $f_{\mathcal{R}}$ to get the cumulative probability function
$F_{\mathcal{R}}$. We then find the age limits $t_0$ and $t_1$ such that the
probabilities

\begin{eqnarray}
P(\sin i > \max\{\sin i_j\}|t_0) &=& \frac{1+\alpha}{2} \quad\mbox{and} \\
P(\sin i > \max\{\sin i_j\}|t_1) &=& \frac{1-\alpha}{2},
\end{eqnarray}

where $\alpha$ is the significance and the probability function is

\begin{equation}
P(\sin i > \max\{\sin i_j\}|t) = 1 - \prod_j
F_{\mathcal{R}}(R_{\mathrm{mod},j}\max\{\sin i_j\}|R_{\mathrm{mod},j},\sigma_j).
\end{equation}

Using the above relations we find the implied ages of $\eta$\,Cha and TWA to be $t_{\eta\,\mathrm{Cha}} = 
13^{+7}_{-6}$\,Myr and $t_{\mathrm{TWA}} = 9^{+8}_{-2}$\,Myr, respectively, where the quoted confidence
interval is of 95\% significance. These ages are slightly higher than, but consistent with, estimates from
literature (6\,Myr for $\eta$\,Cha and 8\,Myr for TWA), indicating that the model radii with 95\% confidence
are good to within $\sim$15\%. 

\section{Summary}
\label{conc}

Rotation and activity are important parameters in the stellar pre-main-sequence evolution, because they trace
changes of interior structure and magnetic fields as well as the dissipation of circumstellar disks. We present 
a spectroscopic study of rotation (measured as $v\sin i$) and chromospheric activity (measured as H$\alpha$ EW) 
for a sample of 74 young stars with spectral types F5--M5 in stellar associations with ages from 6 to 30\,Myr. 
More than half of the objects are still fully convective, while the remaining fraction has already developed a 
radiative core. The analysis is based on an extensive set of multi-epoch high-resolution spectra obtained with the 
6.5\,m Clay Magellan telescope. We achieve a rotational velocity accuracy of $\le 5$\,km\,s$^{-1}$. In the 
following, we summarize our results:

\begin{enumerate}
\item{The range and distribution of H$\alpha$ EWs do not depend significantly on age in the considered age 
range; instead they are a strong function of spectral type (and thus stellar mass). Mid F to early K type
stars have H$\alpha$ in absorption, while most later type objects show emission. Until early K types, the
H$\alpha$ EW are mostly consistent with pure photospheric absorption, while for later spectral types 
chromospheric emission dominates.}
\item{The spectral type at which H$\alpha$ goes into emission in our sample is clearly earlier than in
older clusters Pleiades and Hyades, but later than in very young T~Tauri stars. This indicates a mass
dependence in the lifetime of active chromospheres. Using this as an age criterion, as suggested by 
\citet{1999ASPC..158...63H}, we find that the plausible age of TH is in the range between 10 and 
40\,Myr.}
\item{The chromospheric activity measured in H$\alpha$ clearly declines as a function of age from T~Tauri 
stars (1--5\,Myr) to post T~Tauri stars in our sample (6--30\,Myr) to ZAMS objects (50--100\,Myr).}
\item{Many objects with spectral types later than early K show measurable variability in H$\alpha$ EW
on timescales of weeks and months, which can be attributed to chromospheric processes.}
\item{Most objects in our sample have projected rotational velocities between 5 and 60\,km\,s$^{-1}$. 
Additionally, four ultrafast rotators with $v\sin i$ between 70 and 130\,km\,s$^{-1}$ are seen, all
in BPMG and TH. The maximum and minimum of the $v\sin i$ range decreases between spectral types mid F 
to early K, indicating a dependence of rotation braking on the depth of the convection zone.}
\item{The average rotational evolution between 5 and 30\,Myr is consistent with angular momentum conservation. 
It does not agree well with constant angular velocity i.e.\ `period-locking'. This is the opposite of what
has been observed for ages 1--5\,Myr \citep{2004AJ....127.1029R} and indicates a change in the rotational 
regulation at ages of $\sim$5--10\,Myr, coinciding with the average lifetime of (inner) disks. This may be 
interpreted with a scenario where the rotation is regulated by disk interaction at early ages, while
they are free to spin up after the disks have disappeared.}
\item{By comparing our data with rotational velocities in the Pleiades, we see some evidence for weak
rotational braking on timescales of $\sim 100$\,Myr. This might be an exponential or a Skumanich type rotational 
braking due to stellar winds ($v \propto t^{-1/2}$), which is switched on after the objects have developed 
radiative cores, i.e.\ after $\sim 30$\,Myr.}
\item{The rotation-activity relation, using $L_{\mathrm{H}\alpha} /L_{\mathrm{bol}}$, appears flat and thus
`saturated' in our sample. The maximum level of $L_{\mathrm{H}\alpha} /L_{\mathrm{bol}}$ is $\sim 3 \times 10^{-4}$,
more or less independent of rotational velocity. There is no clear sign of a linear rotation/activity
correlation at low $v\sin i$.} 
\item{The rotation-activity relation of stars with ages $\lesssim 30$\,Myr is similar to 
fully-convective very low mass objects. The flat rotation/activity relation and the weak wind braking seen 
in these two object classes may not be due to `saturation' of a solar-type rotationally driven dynamo, as 
suggested in the standard picture. Instead, the magnetic fields in these young objects are probably generated 
in a fundamentally different way from those in main-sequence stars.}
\item{By comparing our rotational velocities with rotation periods from the literature, we find ages of 
$13^{+7}_{-6}$\,Myr and $9^{+8}_{-2}$\,Myr for $\eta$\,Cha and TWA, respectively, consistent with 
previous estimates from other methods. This agreement indicates that the stellar radii for M dwarfs from models 
by \citet{1998A&A...337..403B} are good within $\sim$15\%.}
\end{enumerate}

\acknowledgments
We thank the anonymous referee for a constructive report. The assistance of the staff at Las Campanas 
Observatory is greatfully acknowledged.

Facilities: \facility{Magellan}

\clearpage

\begin{deluxetable}{lrrrrl}
\tabletypesize{\scriptsize}
\tablecaption{Summary of results for $\eta$\, Cha. \label{etacha}}
\tablewidth{0pt}
\tablehead{
\colhead{Star} & \colhead{H$\alpha$ EW} & \colhead{H$\alpha$ EW $\sigma$} & \colhead{$v\sin i$} 
& \colhead{$v\sin i$ $\sigma$} &
\colhead{Spectral Type \tablenotemark{a}}}
\startdata
$\eta$\,Cha 3                & -1.99  & 0.17  & 10.50  & 0.67 & M3.25\tablenotemark{b}\\
$\eta$\,Cha 4                & -3.40  & 0.61  & 5.96  & 1.00 & K7\tablenotemark{a} \\
$\eta$\,Cha 5                & -8.57  & 4.29  & 8.75  & 1.48 & M4\tablenotemark{b} \\
$\eta$\,Cha 6                & -5.04  & 0.42  & 20.89  & 1.05 & M2\tablenotemark{a} \\
$\eta$\,Cha 10               & -1.15  & 0.28  & $\mathrm{<}$5.0  & 0.93 & K7\tablenotemark{a} \\
\enddata
\tablecomments{The name of the star, H$\alpha$ equivalent width, standard deviation in the 
H$\alpha$ equivalent width, $v\sin i$, and standard deviation in $v\sin i$ are provided.  
A positive H$\alpha$ EW denotes absorption.}
\tablenotetext{a}{\citet{2004ARA&A..42..685Z}}
\tablenotetext{b}{\citet{2004ApJ...609..917L}}
\end{deluxetable}

\begin{deluxetable}{lrrrrl}
\tabletypesize{\scriptsize}
\tablecaption{Summary of results for TWA.}
\tablewidth{0pt}
\tablehead{
\colhead{Star} & \colhead{H$\alpha$ EW} & \colhead{H$\alpha$ EW $\sigma$} & \colhead{$v\sin i$} 
& \colhead{$v\sin i$ $\sigma$} &
\colhead{Spectral Type}}
\startdata
TWA 2a             &    -1.84 & 0.21 & 12.78 & 1.00 & M2e\tablenotemark{b}\\
TWA 3b             &    -6.14 & 0.89 & 12.20 & 6.59 & M3.5\tablenotemark{b}\\
TWA 7              &    -5.82 & 0.85 & $\mathrm{<}$5.0 & 1.59 & M1\tablenotemark{a}\\
TWA 8a             &    -8.00 & 1.42 & $\mathrm{<}$5.0 & 1.15 & M2\tablenotemark{b}\\
TWA 8b             &    -13.3 & 1.84 & 11.20 & 2.52 & $\sim$M5\tablenotemark{c}\\
TWA 9a             &    -2.14 & 0.45 & 11.26 & 0.53 & K5\tablenotemark{b} \\
TWA 9b             &    -4.31 & 0.60 & 8.39 & 0.61 & M1\tablenotemark{b}\\
TWA 10             &    -13.6 & 9.63 & 6.33 & 1.18 & M2.5\tablenotemark{a}\\
TWA 11b            &    -3.45 & 0.62 & 12.11 & 0.93 & M2.5\tablenotemark{b}\\
TWA 12             &    -4.83 & 0.85 & 16.22 & 0.94 & M2\tablenotemark{b}\\
TWA 13a            &    -3.01 & 0.69 & 10.46 & 1.10 & M1e\tablenotemark{b}\\
TWA 13b            &    -3.00 & 0.74 & 10.33 & 1.19 & M2e\tablenotemark{b}\\
TWA 15a            &    -8.81 & 0.51 & 21.33 & 2.16 & M1.15\tablenotemark{b}\\
TWA 15b            &    -8.59 & 1.35 & 32.33 & 1.86 & M2\tablenotemark{b}\\
TWA 18             &    -3.32 & 0.36 & 24.12 & 0.74 & M0.5\tablenotemark{b}\\
TWA 19a            &     2.60 & 0.13 & 28.35 & 0.71 & G5\tablenotemark{b}\\
TWA 21             &    -0.11 & 0.18 & 6.00 & 1.00 & M1\tablenotemark{b}\\
TWA 22             &    -11.5 & 1.95 & 9.67 & 2.37 & M5\tablenotemark{a}\\
TWA 23             &    -2.44 & 0.19 & 14.78 & 1.72 & M1\tablenotemark{a}\\
TWA 24s            &    -0.30 & 0.16 & 13.87 & 1.13 & K3\tablenotemark{a}\\
TWA 25             &    -2.36 & 0.54 & 11.78 & 1.86 & M0\tablenotemark{a}\\
\enddata
\tablecomments{See notes for Table \ref{etacha} for column heading explanation.}
\tablenotetext{a}{\citet{2004ARA&A..42..685Z}}
\tablenotetext{b}{\citet{2004AJ....128.1812D}}
\tablenotetext{c}{Spectral type approximated by visual comparison with other stars in this group.}
\end{deluxetable}

\begin{deluxetable}{lrrrrl}
\tabletypesize{\scriptsize}
\tablecaption{Summary of results for BPMG.}
\tablewidth{0pt}
\tablehead{
\colhead{Star} & \colhead{H$\alpha$ EW} & \colhead{H$\alpha$ EW $\sigma$} & \colhead{$v\sin i$} 
& \colhead{$v\sin i$ $\sigma$} &
\colhead{Spectral Type\tablenotemark{a}}}
\startdata
AO Men             &    -0.56 & 0.12 & 15.96 & 0.69 & K6/7 \\
AU Mic             &    -2.24 & 0.81 & 8.49 & 0.97 & M1 \\
HD 164249          &    5.43 & 0.09 & 22.50 & 1.97 & F5V \\
HD 181327          &    5.09 & 0.05 & 20.83 & 1.33 & F5.5 \\
HD 35850           &    4.31 & 0.18 & 52.00 & 1.00 & F7 \\
PZ Tel             &    2.29 & 0.17 & 77.50 & 2.81 & K0Vp \\
V343 Nor           &    2.22 & 0.15 & 17.00 & 1.00 & K0V \\
GJ 3305            &    -2.15 & 0.27 & 5.30 & 1.03 & M0.5 \\
GJ 799n            &    -10.9 & 1.61 & 10.56 & 2.13 & M4.5e \\
GJ 799s            &    -9.25 & 0.61 & 17.00 & 3.54 & M4.5e \\
HIP 23309          &    -0.77 & 0.13 & 5.77 & 0.73 & M0.5 \\
HIP 23418A         &    -6.55 & 0.00 & 7.67 & 2.08 & M3V \\
HIP 23418B         &    -6.06 & 0.00 & 21.00 & 4.36 & $\sim$M3\tablenotemark{b} \\
HIP 112312         &    -6.61 & 0.14 & 14.00 & 1.73 & M4e \\
HIP 112312B        &    -8.16 & 0.46 & 24.33 & 4.93 & M4e \\
HIP 12545          &    -0.56 & 0.00 & 9.30 & 0.64 & M0 \\
\enddata
\tablecomments{See notes for Table \ref{etacha} for column heading explanation.}
\tablenotetext{a}{\citet{2004ARA&A..42..685Z}}
\tablenotetext{b}{Spectral type approximated by visual comparison with other stars in this group.}
\end{deluxetable}

\begin{deluxetable}{lrrrrl}
\tabletypesize{\scriptsize}
\tablecaption{Summary of results for Tuc-Hor \label{th}}
\tablewidth{0pt}
\tablehead{
\colhead{Star} & \colhead{H$\alpha$ EW} & \colhead{H$\alpha$ EW $\sigma$} & \colhead{$v\sin i$} 
& \colhead{$v\sin i$ $\sigma$} &
\colhead{Spectral Type \tablenotemark{a}}}
\startdata

CD -53544           &  -1.43 & 0.24 & 82.22 & 5.02 & K6Ve \\
CD -60416           &  -0.48 & 0.04 & 10.11 & 0.60 & K3/4 \\
CPD -64120          &  -0.19 & 0.12 & 30.22 & 1.09 & K1Ve \\
GSC 8056-0482       &  -5.31 & 0.44 & 34.22 & 5.61 & M3Ve \\
GSC 8491-1194       &  -4.14 & 0.36 & 12.78 & 0.83 & M3Ve \\
GSC 8497-0995       &  -0.63 & 0.22 & 6.56 & 0.53 & K6Ve \\
HIP 107345          &  -1.41 & 0.15 & 6.44 & 1.33 & M1 \\
HIP 1993           &    -1.01 & 0.11 & 9.50 & 0.55 & M1 \\
HIP 2729           &    -0.68 & 0.26 & 127.5 & 3.94 & K5V \\
HIP 3556           &    -0.79 & 0.02 & $\mathrm{<}$ 5.0 & 0.52 & M3 \\
HD 13183           &    2.70 & 0.13 & 21.00 & 0.71 & G5V \\
HD 13246           &    4.13 & 0.13 & 29.78 & 1.79 & F8V \\
HD 8558            &    2.62 & 0.10 & 12.11 & 0.60 & G6V \\
HD 9054            &    0.76 & 0.11 & $\mathrm{<}$ 5.0 & 0.53 & K1V \\
HIP 105388         &    2.58 & 0.10 & 12.83 & 0.39 & G5V \\
HIP 107947         &    4.26 & 0.05 & 30.42 & 1.38 & F6V \\
HIP 108422         &    1.55 & 0.54 & 139.8 & 9.42 & G8V \\
HIP 1113           &    2.57 & 0.05 & 6.00 & \nodata & G6V \\
HIP 1481           &    4.06 & 0.00 & 22.67 & 0.52 & F8/G0 \\
HIP 16853          &    3.41 & 0.10 & 17.80 & 0.41 & G2V \\
HIP 21632          &    3.02 & 0.00 & 17.53 & 0.64 & G3V \\
HIP 22295          &    3.54 & 0.29 & 41.33 & 2.10 & F7V \\
HIP 30030          &    3.31 & 0.17 & 40.40 & 1.24 & G0 \\
HIP 30034          &    1.30 & 0.26 & 9.73 & 1.71 & K2V \\
HIP 32235          &    2.72 & 0.05 & 10.25 & 0.45 & G6V \\
HIP 33737          &    0.77 & 0.52 & 8.93 & 0.26 & K3V \\
HIP 490            &    3.72 & 0.03 & 14.50 & 0.55 & G0V \\
HIP 9141           &    3.03 & 0.07 & 14.78 & 0.44 & G3/5V \\
TYC 5882-1169      &    0.74 & 0.16 & 6.80 & 0.56 & K3/4 \\
TYC 7065-0n        &    1.45 & 0.02 & 22.33 & 1.03 & K4V \\
TYC 7065-0s        &    1.58 & 0.06 & 14.00 & 0.89 & $\sim$K4\tablenotemark{b}\\
TYC 7600-0         &    1.76 & 0.25 & 18.80 & 0.86 & K1 \\
\enddata
\tablecomments{See notes for Table \ref{etacha} for column heading explanation.}
\tablenotetext{a}{\citet{2004ARA&A..42..685Z}}
\tablenotetext{b}{Spectral type approximated by visual comparison with other stars in this group.}
\end{deluxetable}

\clearpage

\clearpage

\clearpage

\clearpage

\clearpage

\clearpage

\clearpage

\clearpage

\end{document}